\documentclass[twocolumn]{aastex631}

\let\pwiflocal=\iffalse \let\pwifjournal=\iffalse
\usepackage[utf8]{inputenc}
\usepackage{textcomp}
\usepackage{amsmath,amssymb,gensymb,times,graphicx,morefloats,color}
\usepackage{afterpage, bm}
\usepackage{natbib,hyperref}
\usepackage[outdir=./]{epstopdf}
\usepackage{mathrsfs}
\usepackage{mathtools}
\usepackage{url}
\usepackage{rotating}
\newcommand{\teff}{\ensuremath{T_{\text{eff}}}}
\newcommand{\logg}{log$(g)$}

\newcommand{\rsun}{$R_\odot$}

\newcommand{\msun}{$M_\odot$}

\newcommand{\funit}{erg s$^{-1}$ cm$^{-2}$ \AA$^{-1}$}
\newcommand{\vsini}{$v$sin$i$}

\newcommand{\mdot}{$\dot{M}$}

\newcommand{\kms}{km s$^{-1}$}
\newcommand{\gaia}{{\it Gaia}}
\newcommand{\hst}{{\it HST}}

\newcommand{\tess}{{\it TESS}}

\newcommand{\co}{$^{12}{\mathrm{CO}} \ J=2-1$}

\graphicspath{{./}{}}

\shortauthors{Tofflemire et al.}
\shorttitle{Binary-Disk Alignment in FO Tau}

\begin{document}

\title{Sites of Planet Formation in Binary Systems. I. 
Evidence for Disk-Orbit Alignment in the Close Binary FO Tau}

\correspondingauthor{Benjamin M.\ Tofflemire}
\email{tofflemire@utexas.edu}

\author[0000-0003-2053-0749]{Benjamin M.\ Tofflemire}
\altaffiliation{51 Pegasi b Fellow}
\affiliation{Department of Astronomy, The University of Texas at Austin, Austin, TX 78712, USA}

\author[0000-0001-7998-226X]{Lisa Prato}
\affiliation{Lowell Observatory, 1400 West Mars Hill Road, Flagstaff, AZ 86001 USA}

\author[0000-0001-9811-568X]{Adam L.\ Kraus}
\affiliation{Department of Astronomy, The University of Texas at Austin, Austin, TX 78712, USA}

\author[0000-0003-3172-6763]{Dominique Segura-Cox}
\altaffiliation{NSF AAPF Fellow}
\affiliation{Department of Astronomy, The University of Texas at Austin, Austin, TX 78712, USA}

\author[0000-0001-5415-9189]{G.\ H.\ Schaefer}
\affiliation{The CHARA Array of Georgia State University, Mount Wilson Observatory, Mount Wilson, CA 91023, USA}

\author[0000-0001-9674-1564]{Rachel Akeson}
\affiliation{NASA Exoplanet Science Institute, IPAC/Caltech, Pasadena, CA, 91125, USA}

\author[0000-0003-2253-2270]{Sean Andrews}
\affiliation{Center for Astrophysics $|$ Harvard \& Smithsonian, 60 Garden Street, Cambridge, MA 02138, USA}

\author[0000-0002-4625-7333]{Eric L.\ N.\ Jensen}
\affiliation{Dept. of Physics \& Astronomy, Swarthmore College, 500 College Ave., Swarthmore,
PA 19081, USA}

\author[0000-0002-8828-6386]{Christopher M. Johns-Krull}
\affiliation{Physics \& Astronomy Department, Rice University, 6100 Main St., Houston, TX 77005, USA}

\author[0000-0002-9849-5886]{J.\ J.\ Zanazzi}
\altaffiliation{51 Pegasi b Fellow}
\affiliation{Astronomy Department, Theoretical Astrophysics Center, and Center for Integrative Planetary Science, University of California, Berkeley, \\
Berkeley, CA 94720, USA}

\author{M.\ Simon}
\affiliation{Department of Physics and Astronomy, Stony Brook University, Stony Brook, NY 11794, USA}

\begin{abstract}

Close binary systems present challenges to planet formation. As binary separations decrease, so too do the occurrence rates of protoplanetary disks in young systems and planets in mature systems. For systems that do retain disks, their disk masses and sizes are altered by the presence of the binary companion. Through the study of protoplanetary disks in binary systems with known orbital parameters, we seek to determine the properties that promote disk retention and, therefore, planet formation. In this work, we characterize the young binary-disk system, FO Tau. We determine the first full orbital solution for the system, finding masses of $0.35^{+0.06}_{-0.05}\ M_\odot$ and $0.34\pm0.05\ M_\odot$ for the stellar components, a semi-major axis of $22(^{+2}_{-1})$ AU, and an eccentricity of $0.21(^{+0.04}_{-0.03})$. With long-baseline ALMA interferometry, we detect 1.3mm continuum and $^{12}{\mathrm{CO}} \ (J=2-1)$ line emission toward each of the binary components; no circumbinary emission is detected. The protoplanetary disks are compact, consistent with being truncated by the binary orbit. The dust disks are unresolved in the image plane and the more extended gas disks are only marginally resolved. Fitting the continuum and CO visibilities, we determine the inclination of each disk, finding evidence for alignment of the disk and binary orbital planes. This study is the first of its kind linking the properties of circumstellar protoplanetary disks to a precisely known binary orbit. In the case of FO Tau, we find a dynamically placid environment (coplanar, low eccentricity), which may foster its potential for planet formation.

\end{abstract}

\section{Introduction} 

Binary stars can profoundly shape the dynamics of the circumstellar environment. In the epoch of planet formation, their orbital motion can alter the distribution and internal kinematics of protoplanetary disks. Stellar binarity is generally expected to impede the formation of planetary systems \citep{Nelson2000,Zsometal2011,Mulleretal2012,Picogna&Marzari2013,Jordanetal2021,Zagariaetal2023}, but the detailed orbital properties and formation pathways of binary systems that foster planet formation remain largely unknown. Sub-mm observations at high angular resolution, possible with the ALMA observatory, facilitate the investigation of the binary-disk interaction and its effect on planet formation.  

The impact binarity has on the planet forming reservoir is, to first order, dependent on the binary separation. Population studies of (sub-)mm continuum emission (tracing $\sim$mm size dust grains) in young binaries find that beyond separations of a few 100 AU, binary disks are indistinguishable from those orbiting single stars \citep{Akesonetal2019}. Interior to this value, circumstellar disks become less massive \citep{Jensenetal1996,Harrisetal2012,Barenfeldetal2019} and/or smaller \citep{Coxetal2017,Zurloetal2020,Zurloetal2021} as the binary projected separation decreases. Large dust grains can be biased tracers of disk properties due to processes like radial drift \citep{Weidenschilling1977,Ansdelletal2018}, but the trend in disk radii follows the general predictions of binary-disk truncation theory \citep[e.g.,][]{Artymowicz&Lubow1994}, in that systems with smaller separations have smaller disks. For completeness, we note that disk material can also reside in circumbinary orientations \citep[e.g.,][]{Czekalaetal2019}, but these architectures become less common at larger binary separations \citep[e.g.,][]{Akeson&Jensen2014,Akesonetal2019}. 

Sub-mm observations combining dust and gas provide a more complete description of the binary-disk interaction. For a sample of Taurus binaries, \citet{Manaraetal2019} find that dust disks are smaller and have sharper cutoffs in their radial profiles than single stars. Including gas observations for this sample, \citet{Rotaetal2022} find gas disk radii that are consistent with dynamical truncation theory, assuming modest orbital eccentricities (since the actual eccentricities are not uniformly known). Additionally, they find the average ratio of the gas radius to the dust radius is larger for binaries ($\sim4$) than single stars \citep[$\sim3$;][]{Ansdelletal2018,Kurtovicetal2021,Sanchisetal2021}. Models of radial drift in truncated disks have been shown reproduce this difference \citep{Zagariaetal2021a,Zagariaetal2021b} without invoking any additional dynamical effects \citep[e.g., dynamical stirring;][]{Quintanaetal2007}.

Beyond the characterization of disk-bearing binary systems, the absence of disk material is also relevant. The occurrence rate of disks in binary systems is known to decrease with decreasing binary separation \citep{Ciezaetal2009,Krausetal2012,Cheetham2015}. It is unclear whether this trend is driven by the rapid exhaustion of a truncated disk, a dependence of disk survival on specific orbital parameters (e.g., high eccentricity, mutual disk-orbit inclination), or a product of the binary formation mechanism \citep[e.g., disk fragmentation or core fragmentation, or either with subsequent migration, see][]{Offneretal2022}.

Inferring the consequence these observational results carry for planet formation is challenging. Empirically, the occurrence rate of transiting planets in field-age binary systems declines as the binary separation decreases \citep{Wangetal2014,Krausetal2016,Moe&Kratter2021}. The trend closely follows that of the binary-disk occurrence rate, signaling an intuitive link between disk retention and planet formation. Yet, for the significant minority of binary systems with separations less than 50 AU that retain long-lived disks and form planets, it is unclear what properties these systems have that allow them to do so. Emerging results from the binary transiting-planet population suggest a preference for mutual alignment of the planet and binary orbital planes \citep{Dupuyetal2022,Behmardetal2022,Christianetal2022,Lesteretal2023}.

A critical obstacle toward developing a predictive theory of planet formation in binary systems is linking protoplanetary disk properties we can now measure with ALMA to the binary orbital and detailed stellar parameters. The disk studies above have had little more than the projected binary separation to interpret their observations. This limitation is rooted in the difficulty of determining orbital solutions for binaries with separations of a few 10s of AU. At the distances of the closest star-forming regions ($\sim140$ AU), decades of high-angular-resolution imaging are required to measure full orbits or sufficient arcs. While challenging, these are the prime targets where the binary-disk interaction is pronounced and where binary separations are large enough to allow disk sizes, in principle, that can be resolved with current observatories. 

To connect the properties of protoplanetary disks to those of the binary orbit, we have observed 1.3 mm continuum and the \co\ transition using the ALMA observatory for a sample of eight disk-hosting binary systems with known orbital solutions. Observing in extended array configurations, we seek to characterize the binary-disk interaction by resolving the location and distribution of circumstellar material. In this study, we present our analysis of FO Tau, a $\sim150$ milliarcsecond (mas) separation binary ($\sim$20 AU). FO Tau has the largest projected separation in our sample, making it a convenient target to demonstrate our methodology. A comprehensive analysis of the full sample will follow. We present the first full orbital solution for the system, combining astrometric and radial velocity (RV) measurements, and forward-model the ALMA visibilities to determine the obliquity of the individual circumstellar disks with the binary orbital plane. This work provides one of the first analyses linking stellar and protoplanetary dynamics marking a critical step toward understanding planet formation in the binary environment.

\section{Observations \& Data Reduction}

\subsection{ALMA}

\subsubsection{Observations}

Our ALMA observations of FO Tau took place in Cycle 7 (Project Code 2019.1.01739.S) using Band 6 receivers in dual polarization mode. Three spectral windows were placed to sample continuum at 231.6, 234.0, and 245.9 GHz, each with a bandwidth of 1.875 GHz sampled at 31.25 MHz resolution. A fourth spectral window was centered on the \co\ transition (230.538 GHz) at the source's heliocentric velocity ($\sim$16 km/s; \citealt{Kounkeletal2019}). This CO window covered 938 MHz with a two-channel velocity resolution of 488 kHz (0.64 \kms). This setting was chosen to yield a velocity resolution required to resolve a Keplerian rotation profile while maintaining a broad bandwidth for continuum sensitivity. 

To achieve high angular resolution while maintaining sensitivity to extended emission, observations were made in compact and extended array configurations. The compact configuration had baselines between 14 m and 3.6 km, corresponding to an angular resolution and maximum recoverable scale of 0.1$\arcsec$ and 2.1$\arcsec$, respectively, in Band 6 (roughly the C6 configuration). The extended configuration baselines spanned 40 m to 11.6 km, reaching an angular resolution of 0.034$\arcsec$ and maximum recoverable scale of 0.83$\arcsec$ (roughly the C8 configuration). Observations in the compact and extended configurations took place on 2021-07-18--11:17 and 2021-08-22--09:15 with 6 and 50 minute on-source times, respectively, following the standard calibrator observing sequence. The angular scales these observations correspond to a spatial resolution of $\sim$5 AU and a maximum recoverable extent of emission of $\sim$280 AU at FO Tau's distance ($d=135\pm4$ pc). This sensitivity range covers the physical scales of interest for our analysis, specifically probing both circumstellar and circumbinary emission.

\subsubsection{Calibration and Imaging}

Extended and compact configurations were calibrated separately by the standard ALMA pipeline ({\tt CASA v6.1.1.15}; \citealt{McMullinetal2007}) and provided as a fully-calibrated measurement set by the North American ALMA Science Center. The calibration includes data flagging, phase corrections from water vapor radiometer measurements, bandpass and amplitude calibration, and a temporal gain correction. From the pipeline-calibrated data we followed the post-processing procedures developed for the DSHARP program, which focus on combining array configurations and self-calibration. A full description of this approach with accompanying scripts can be found in \citet{Andrewsetal2018} and the program's data release web page\footnote{https://almascience.eso.org/almadata/lp/DSHARP/}. We provide an outline of our procedure below using {\tt CASA v6.5.1.23}. 

First, we created a continuum data set for each measurement set (compact, extended), removing channels within $\pm$20 \kms\ of the \co\ transition at the FO Tau systemic velocity. We imaged each data set using {\tt tclean} (Briggs weighting, {\tt robust} = 0) to test the astrometric alignment and flux scaling. A shift was made to align the phase center of each measurement set on the FO Tau A continuum source. The flux scaling between the data sets was consistent within the uncertainty (flux ratio of $0.999\pm0.006$). 

Next we began an iterative phase-only self-calibration process on the compact configuration. We imaged the compact configuration with a broad cleaning mask that encompassed both stellar components, which created a source model we used to calibrate the visibilities against. An annulus outside this mask was used to measure the signal-to-noise ratio (SNR). We repeated this process, reducing the solution interval until one of the following conditions was met: the SNR does not improve from the previous iteration, the number of flagged solutions for any time interval exceeds 20\%, or we reach the native integration step (6 s). Solution intervals of the total duration and of 100 s improved the SNR from 52 to 84 before plateauing at 60 s. One iteration of amplitude-only self calibration over the full integration was then performed, which improved the SNR to 85. 

We then combined the extended and self-calibrated compact measurement sets and repeated this process. Phase-only self-calibration was performed for the full integration, 1500 s, 1000 s, and 600 s, bringing the SNR from to 60 to 150 before leveling out. An iteration of amplitude-only self-calibration did not significantly improve the data quality, leaving the last iteration of phase-only self-calibration as our fiducial continuum data set. Final continuum imaging was performed with interactive masking using Briggs weighting ({\tt robust} = 0) with a three sigma cleaning threshold. This resulted in a synthesized continuum beam of $45\times22$ mas with a position angle of $22^\circ$. The root-mean-square deviation (RMS) of the final image is 0.03 mJy/beam. This image and its analysis are presented in Section \ref{continuum}.

From the initial pipeline-calibrated measurement sets, we also created \co\ data cubes for the compact and extended configurations, including channels within $\pm$20 \kms\ from FO Tau's systemic velocity. The same self-calibration steps above were applied, followed by continuum subtraction. Here we adopted the 2-channel spacing of 0.635 \kms. The resultant combined cube was imaged with interactive masking using Briggs {\tt robust} = 1 weighting (which is larger than the continuum imaging due to a lower average SNR in the CO cube). The synthesized CO beam is $79\times36$ mas with a position angle of $25^\circ$. The CO channel maps have a RMS values of 0.9 mJy/beam and are cleaned to a depth of three sigma. These maps are presented in Section \ref{co} and \ref{disk_fit}.

\subsection{Keck -- NIRSPEC behind AO}
\label{nirspec_obs}

FO Tau was observed on UT 2010 December 12 (UTC 09:32:18.66) at the Keck II 10-meter telescope using NIRSPEC \citep{McLeanetal2000} behind the adaptive optics (AO) system. The spectrograph slit position angle was set to 230 degrees in order to align both components of the close binary on the slit simultaneously; the 2-pixel slit yielded a resolution of $\sim$30,000. The AO system re-images the instrumental dimensions onto the detector reduced by a factor of $\sim$11, thus the post-AO slit width$\times$length used was
$0\farcs027\times2\farcs26$. At the time of observations, the airmass was 1.01 and the natural seeing was $0\farcs5$, facilitating robust correction from the AO system. The AO frame rate for FO Tau was 149 Hz and wavefront sensor counts were around 265.

We observed FO Tau in the $H$-band with the N5 filter, with 1.555$\mu$m near the middle of the central order (49). This spectral region is rich in atomic and molecular lines suitable for the measurement of fundamental atmospheric and other properties in low-mass stars spanning effective temperatures of 3000--5000 K. Keck's cold, high altitude, and low humidity site on Mauna Kea ensures that no significant telluric absorption is present in this near-infrared (NIR) spectral range; thus the data shown in Figure \ref{fig:nirspec} have not been divided by an A0 star.

Internal comparison lamp lines were used to obtain a second order dispersion solution; median filtered stacks of ten internal flat lamp and dark frame exposures yielded a master flat and dark. A continuum SNR of $\sim$130 was achieved with six 300 s exposures of the target binary, taken in nodded pairs along the slit to facilitate background subtraction. Differenced nod pairs were divided by the final dark-subtracted flat. We used the REDSPEC reduction package \citep{Kimetal2015} for the array rectification and spectral extraction.

\subsection{High-Angular Resolution Imaging}
\label{har-obs}

We obtained three new AO observations of FO Tau using the NIRC2 camera at the Keck Observatory \citep{Wizinowich2000} on UT 2016 Oct 20, 2019 Jan 20, and 2022 Oct 19. On each night, we obtained sets of 12 images dithered across the detector in the Hcont and Kcont filters. Each image consisted of 1.0 sec exposures with 10 co-added frames. The images were flat-fielded using dark-subtracted, median-combined dome flats. Pairs of dithered images were subtracted to remove the sky background. We observed a point spread function (PSF) reference star using the same AO frame rate either before or after the observations of FO Tau. On the first two nights we used DN Tau as a PSF reference, while DH Tau was used on the third night. Our three observations were take at airmasses of 1.07, 1.12, and 1.02, respectively. In each case the AO system easily separated two sources, producing a clear airy ring for each component.

We fit a binary model to the images of FO Tau using the PSF fitting technique described by \citet{Schaeferetal2014}. The procedure uses the PSF star to construct a binary model and searches through a grid of separations and flux ratios to solve for the best fitting binary separation, position angle, and flux ratio. On UT 2016 Oct 20, the shape of the AO corrected images varied over time because of an oscillation problem with the Keck II telescope (C. Alvarez; private communication). To account for the changing PSF shape between individual images, we created an effective PSF for each image using the two components in FO Tau following the approach described by \citet{Schaefer2018}, instead of using the observed PSF reference star to model the binary. We corrected the binary positions using the geometric distortion solutions published by \citet{Service2016} and applied a plate scale of 9.971 $\pm$ 0.004 mas~pixel$^{-1}$ and subtracted $0\fdg262 \pm 0\fdg020$ from the measured position angles. The positions are averaged over the measurements from individual images, and uncertainties are computed from the standard deviation. 

We also obtained AO observations of FO Tau with NIRC2 in the $J$, Hcont, and Kcont filters on UT 2009 Nov 21. We obtained two images in a single dither position, each consisting of 40 co-added exposures of 0.5 sec each. The seeing was better than average ($\sim$0.4\arcsec) and the system was observed at moderate airmass ($\sim$1.55), yielding diffraction-limited PSFs even at $J$ band. The images were processed and analyzed with PSF-fitting photometry following the methods described by \citet{Krausetal2016}. To briefly recap, the relative astrometry and photometry of the FO Tau binary was fit within each image by testing the $\chi^2$ goodness-of-fit for a library of potential empirical templates (encompassing the 1000 images of single stars that were taken closest to that date) using an initial estimate of the photometry/astrometry, then the best empirical template was used to optimize the fit of the photometry and astrometry, and we iterated between the two stages until the same empirical template produced the same best-fit values. The values for the two images were then averaged, with the nominal two-value RMS being adopted as the uncertainty because it matches the typical RMS ($\sim$0.005 mag) for contrast measurements of other bright, similar-contrast binaries separated by $\sim$3 $\lambda/D$.

Table \ref{tab:AO} presents the astrometric measurements from the literature and this work, which are included in our fit to the FO Tau binary orbit. The flux ratios of FO Tau B relative to A measured from our Keck AO observations are given in Table \ref{tab:flux}. The AO observations are available for further inspection on the Keck Archive\footnote{\url{https://koa.ipac.caltech.edu/cgi-bin/KOA/nph-KOAlogin}}.

\begin{deluxetable*}{l c c c c}
\tablecaption{High-Angular Resolution Orbit Monitoring
\label{tab:AO}}
\tablewidth{0pt}
\tabletypesize{\footnotesize}
\tablecolumns{5}
\phd
\tablehead{
  \colhead{Date (year)} &
  \colhead{$\rho$ (\arcsec)} &
  \colhead{P.A. ($^\circ$)} &
  \colhead{Provenance} &
  \colhead{Reference}
}
\startdata
1991.7139 & 0.165$\pm$0.005   & 180.0$\pm$4.0  & Calar Alto 3.5m-Speckle & 1 \\
1991.7933 & 0.161$\pm$0.001   & 181.7$\pm$0.6  & Palomar Hale 5m-Speckle & 2 \\
1993.7591 & 0.183$\pm$0.005   & 182.0$\pm$0.9  & Calar Alto 3.5m-Speckle & 3 \\
1994.7995 & 0.153$\pm$0.002   & 190.6$\pm$0.4  & Palomar Hale 5m-Speckle & 2 \\
1994.9500 & 0.159$\pm$0.004   & 189.7$\pm$0.9  & Calar Alto 3.5m-Speckle & 3 \\
1994.9637 & 0.154$\pm$0.002   & 191.2$\pm$0.4  & Palomar Hale 5m-Speckle & 2 \\
1995.7687 & 0.156$\pm$0.004   & 193.1$\pm$1.3  & Calar Alto 3.5m-Speckle & 3 \\
1996.7379 & 0.152$\pm$0.004   & 194.3$\pm$0.9  & Calar Alto 3.5m-Speckle & 3 \\
1996.9103 & 0.143$\pm$0.004   & 200.0$\pm$1.1  & Calar Alto 3.5m-Speckle & 3 \\
1996.9268 & 0.145$\pm$0.007   & 193.7$\pm$1.0  & IRTF-Speckle            & 4 \\
1996.9295 & 0.160$\pm$0.001   & 188.4$\pm$5.9  & IRTF-Speckle            & 4 \\
1997.1759 & 0.153$\pm$0.003   & 194.7$\pm$0.4  & \hst-WFPC2              & 4 \\
1997.8741 & 0.150$\pm$0.005   & 198.9$\pm$0.8  & Calar Alto 3.5m-Speckle & 3 \\
1999.79   & 0.146$\pm$0.001   & 203.8$\pm$4.1  & \hst-STIS               & 5 \\
2001.107  & 0.145$\pm$0.004   & 207.6$\pm$0.6  & Calar Alto 3.5m-Speckle & 6 \\
2001.838  & 0.149$\pm$0.004   & 206.7$\pm$1.3  & Calar Alto 3.5m-Speckle & 6 \\
2009.888  & 0.138$\pm$0.002   & 234.9$\pm$0.6  & Keck-NIRC2 AO           & 7 \\
2011.7780 & 0.1372$\pm$0.0002 & 241.1$\pm$0.1  & Keck-NIRC2 AO           & 8 \\
2013.0729 & 0.134$\pm$0.002   & 245.9$\pm$0.8  & Keck-NIRC2 AO           & 8 \\
2015.7151 & 0.124$\pm$0.012   & 253.2$\pm$3.7  & ALMA                    & 9 \\
2016.8024 & 0.1375$\pm$0.0018 & 258.1$\pm$0.7  & Keck-NIRC2 AO           & This Study \\
2019.0523 & 0.1369$\pm$0.0022 & 266.0$\pm$0.9  & Keck-NIRC2 AO           & This Study \\
2021.5437 & 0.1370$\pm$0.0005  & 274.7$\pm$0.2  & ALMA                    & This Study \\
2022.7982 & 0.1359$\pm$0.0010  & 278.3$\pm$0.2  & Keck-NIRC2 AO           & This Study \\
\enddata
\tablecomments{
(1) \cite{Leinertetal1993};
(2) \cite{Ghezetal1995};
(3) \cite{Woitasetal2001};
(4) \cite{WhiteGhez2001};
(5) \cite{HartiganKenyon2003};
(6) \cite{Tamazianetal2002};
(7) Keck Archive D N2.20091121.2475 (PI - Hillenbrand);
(8) \cite{Schaeferetal2014};
(9) \cite{Akesonetal2019}
}
\end{deluxetable*}

\begin{deluxetable}{l c c c}
\tablecaption{Flux ratios for FO Tau from Keck adaptive optics observations
\label{tab:flux}}
\tablewidth{0pt}
\tabletypesize{\footnotesize}
\tablecolumns{4}
\phd
\tablehead{
  \colhead{Date (year)} &
  \colhead{} &
  \colhead{Flux Ratio ($f_B/f_A$)} &
  \colhead{} \\
  \colhead{} &
  \colhead{$J$} &
  \colhead{$H_{cont}$} &
  \colhead{$K_{cont}$} 
}
\startdata
2009.888  & $0.932 \pm 0.005$ & $0.885 \pm 0.003$ & $0.821 \pm 0.003$ \\
2016.8024 & ...               & $0.880 \pm 0.005$ & $0.810 \pm 0.004$ \\
2019.0523 & ...               & $0.862 \pm 0.054$ & $0.821 \pm 0.010$ \\
2022.7982 & ...               & $0.891 \pm 0.006$ & $0.793 \pm 0.006$ \\
\enddata
\end{deluxetable}

\subsection{Time-Series Photometry}

FO Tau was observed by the Transiting Exoplanet Survey Satellite (\tess; \citealt{Rickeretal2015}) in Sectors 19, 43, and 44 with a 2 minute cadence. The photometry was reduced with the SPOC pipeline \citep{Jenkins2015,Jenkinsetal2016}. Our analysis makes use of the simple-aperture photometry (SAP) and the pre-search data conditioning simple aperture
photometry \citep[PDCSAP;][]{SmithKepler2012,StumpeKepler2012,StumpeMultiscale2014} light curves. All of the \tess\ data used in this paper can be found in MAST: \dataset[10.17909/zpka-v291]{http://dx.doi.org/10.17909/zpka-v291}.

\subsection{Literature Photometry \& Astrometry}

Our analysis makes use of photometric measurements from the literature and astrometry from \gaia\ DR3. The relevant measurements are included in Table \ref{tab:char}. 

Close binaries ($\rho<1\arcsec$) are known to complicate the \gaia\ astrometric pipeline, realized most notably through the Renormalized Unit Weight Error (RUWE; \citealt{Lindegrenetal2018}) parameter. RUWE values above $\sim$1.2 have been shown to indicate the presence of unresolved companions in main sequence stars \citep{Rizzutoetal2018,Belokurovetal2020,Brysonetal2020a}. In disk bearing stars, this threshold is closer to 2.5 \citep{Fittonetal2022}. FO Tau's RUWE is 11.499, in line with its known binarity. 

The degree to which this elevated uncertainty introduces additional random or systematic errors is unclear. Most relevant for the current analysis is the parallax and the distance derived from it. The \citet{Bailer-Jonesetal2021} geometric distance is $135^{+4}_{-3}$ pc, while clustering analyses of Taurus members places FO Tau's sub-group center at $\sim$130 pc with a radial dispersion of a few pcs \citep{Esplin&Luhman2019,Krolikowskietal2021}. Given that these values are in fair agreement, we adopt the geometric distance in our analysis.

\section{Stellar and Binary Orbital Parameters}

In this section we describe our measurements and modeling of the stellar and orbital parameters of the FO Tau binary. A summary of our results is provided in Table \ref{tab:char}. 

\subsection{Stellar Characterization}
\label{stellar}

\subsubsection{Modeling High-Resolution NIR Spectra}
\label{model_spec}

\begin{figure}[t!]
\begin{center}
\includegraphics[width=0.48\textwidth]{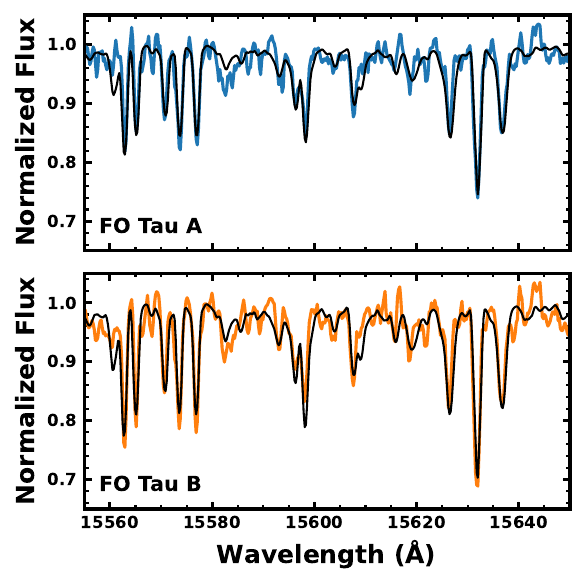}
\caption{Spatially resolved spectra of FO Tau A (top) and B (bottom). The continuum normalized spectra are presented in colored lines and the best-fit model is overlaid in black.}
\label{fig:nirspec}
\end{center}
\end{figure}

The $H$-band is an information rich spectral region for cool stars, containing diagnostics for the typical stellar parameters (\teff, \vsini, \logg), as well as the magnetic field strength \citep[e.g.][]{Hanetal2023}. Figure \ref{fig:nirspec} presents a telluric-free region of the FO Tau spectra. The two spectra are very similar. The most readily apparent difference is the line depths, which are shallower for FO Tau A owing to a high contribution from veiling. 

To model these spectra, we develop a grid of $H$-band spectra using the Spectroscopy Made Easy spectral synthesis code of \citet{Valenti&Piskunov1996} along with the NextGen atmosphere models of \citet{Allard&Hauschildt1995}. Our grid covers the range of \teff, \vsini, magnetic field strength, surface gravity, and veiling expected for young, low-mass stars. The spectra are synthesized using laboratory atomic transition data from the Vienna Atomic Line Database \citep{Piskunovetal1995,Ryabchikovaetal2015} following the procedure of \citet{Johns-Krulletal1999,Johns-Krull2007}. The synthetic spectra are then calibrated against the Solar spectrum \citep{Livingston&Wallace1991} and the spectrum of 61 Cyg B \citep{Wallace&Hinkle1996}. Following an iterative procedure, oscillator strengths and van der Waals broadening constants are adjusted in order to obtain the optimum match of the synthesized spectra to the observed standard spectra. Using initial values already spectroscopically estimated, we run our spectra through this grid to obtain estimates for the complete set of desired properties for the young binary component stars in our sample. The best fitting models are shown as black lines in Figure \ref{fig:nirspec}. The best fit values are compiled in Table \ref{tab:char}. 

The primary errors quoted in Table \ref{tab:char} are internal fitting uncertainties and do not include systematic or additional astrophysical sources of error. These sources of uncertainty are particularly relevant for the \teff\ determination in this work. Systematic \teff\ uncertainties can arise from the choice of the model grid. One clear example comes from \citet{Lopez-Valdiviaetal2021}, where including the magnetic field as a fit parameter increases the measured \teff\ by $\sim$ 40 K compared to a non-magnetic model (for stars with similar magnetic field strengths to FO Tau). A wavelength dependence in the \teff is also observed, attributed to the chromatic dependence of the spot contribution \citep[e.g.,][]{Gully-Santiagoetal2017}. Astrophysically, young stars also have high spot covering fractions \citep[e.g.,][]{Fangetal2016,Gully-Santiagoetal2017,Caoetal2022} that vary with the stellar rotation. In a sample of about a dozen young stars in Taurus, S.-Y. Tang et al.\ (2024, in prep, private communication; \citealt{Tangetal2023AAS}) find changes in \teff\ on the order of $\sim100$ K, based on the rotational phase from $H$-band line equivalent width ratios. Given these effects, a \teff\ uncertainty of 100 K is a more conservative value, which we include as the parenthetical in Table \ref{tab:char}.

\

\subsubsection{Modeling the Spectral Energy Distribution}
\label{model_flux}

In this section we constrain the stellar masses, specifically the mass ratio, $q$, and total mass, $M_{\rm total}$, to place informed priors on the binary orbit fit in Section \ref{orbit}. With only $\sim$one quarter of the orbit traced with astrometry, the current observational baseline does not precisely constrain the stellar masses on its own. To establish these priors, we fit model spectra to measurements of the stellar photospheric flux to infer stellar effective temperatures and radii. With these, we derive mass priors from the HR diagram using stellar evolution models. 

\begin{figure*}[t!]
\begin{center}
\includegraphics[width=0.98\textwidth]{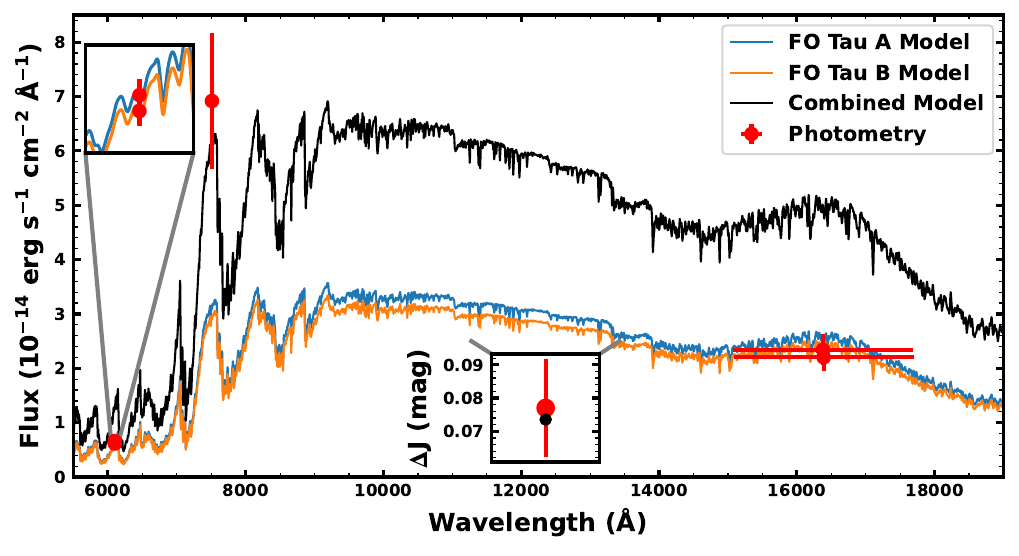}
\caption{Two-component SED fit to FO Tau photospheric flux measurements. The primary and secondary best-fit BT-Settl CFIST spectra are shown in blue and orange, respectively, with a combined spectrum in black. Model spectra have been smoothed to a resolution of $R\sim3000$ for clarity. Red points present measurements of the photospheric flux as they are fit. Horizontal bars represent the FWHM of the filter or spectral region considered in the measurement. Leftmost are spatially resolved \hst-STIS flux measurements at 6100 \AA\ from \citet{HartiganKenyon2003}, highlighted in the top-left subplot. The point at 7510 \AA\ is the unresolved flux from \citet{HerczegHillenbrand2014}. The lower subplot presents the model agreement with the $J$-band contrast determined in this work. The $\Delta J$ magnitude of the best-fit model is shown as the black point. Grey lines highlight the $J$-band spectral region. Rightmost are the $H$-band photospheric fluxes determined in this work. (Photometry values are provided in Table \ref{tab:char}.)}
\label{fig:sed}
\end{center}
\end{figure*}

Both FO Tau components hosts disks (NIR excess), are actively accreting (variable optical/UV excess), and have significant extinction, which complicates a direct fit of the broadband spectral energy distribution. With the current data set, we can make these corrections for these effects to compute an $H$-band photospheric flux for each component. Beginning with the unresolved 2MASS $H$-band apparent magnitude \citep{Skrutskie2006}, we correct for extinction by computing the $H$-band extinction ($A_H$) adopting $A_H/A_V=0.19(\pm0.03)$ (\citealt{Fitzpatrick1999} extinction law, updated in the NIR by \citealt{Indebetouwetal2005}) and the average visual extinction measured by \citet{Hartiganetal1995} and \citet{HerczegHillenbrand2014} from flux-calibrated, low-resolution optical spectra ($A_V=1.98\pm0.20$). The average (error weighted) flux ratio from the spatially-resolved NIRC2 $H$-band photometry (Table \ref{tab:flux}) is then used to split the corresponding flux between the two components, where we have propagated the flux ratio RMS as the uncertainty. Finally, the veiling measurements from our spatially-resolved, $H$-band spectra (Section \ref{model_spec}, Table \ref{tab:char}) are used to remove the contribution of each disk. A 30\% error on the veiling is propagated following the average level of H-band veiling variability found in \citet{Sousaetal2023}. The resulting photospheric $H$-band flux is provided in Table \ref{tab:char}. 

To supplement these values, we also adopt literature measurements of photospheric fluxes that have been corrected for extinction and accretion, namely, spatially-resolved fluxes at 6100 \AA\ from \hst-STIS spectra \citep{HartiganKenyon2003}, and a spatially unresolved flux at 7510 \AA\ from ground-based spectra \citep{HerczegHillenbrand2014}. The latter was modeled as a single star, but that is incidental for our use here given the similarity of the primary and secondary (Section \ref{model_spec}). Both studies measure consistent V-band extinctions, 1.9 and 2.05, respectively, but given the separate analyses, we inflate the uncertainties on these measurements, corresponding to $\pm0.2$ magnitudes of extinction using the \citet{Cardellietal1989} law, which are added in quadrature to the quoted errors. These values and the inflated errors are presented in Table \ref{tab:char}. 

Additionally, we include a spatially resolved $J$-band contrast from Keck-NIRC2 (Section \ref{har-obs}). This observation is not corrected for accretion or disk veiling, but we include it in our analysis because the $J$-band falls at the minimum of each process's contribution. An additional uncertainty based on the $H$-band contrast variability is added in quadrature. This measurements and its inflated uncertainty are presented in Table \ref{tab:char}.

We fit these measurements (two sets of resolved flux values, an unresolved flux, and a magnitude contrast) with synthetic BT-Settl (CFIST) atmospheric models \citep{Allard2013}, interpolating over effective temperature and \logg\ values between 3.0 and 4.5. The model has six parameters: a \teff\ and radius for each component, a shared \logg, and a distance. Gaussian priors are placed on the effective temperatures, informed by the $H$-band spectral fit including the inflated uncertainty (Section \ref{model_spec}, Table \ref{tab:char}), and on the distance (Table \ref{tab:char}). The fit is made within a Markov Chain Monte Carlo (MCMC) framework using {\tt emcee} \citep{Foreman-Mackeyetal2013} with 28 walkers. Fit convergence is assessed on the fly by measuring the chain auto-correlation. The fit ends once the chain auto-correlation changes by less than 5\% or the number of steps exceeds the auto-correlation time by a factor of 50. The first five auto-correlation times are discarded as burn in. 

Figure \ref{fig:sed} presents the result of our best-fit model compared to the photospheric flux measurements. The fit favors lower effective temperatures than those measured in the $H$-band, highlighting, perhaps, some of the model and/or wavelength dependence of the measurement described in Section \ref{model_spec}. We find effective temperatures of $3370^{+50}_{-60}$ K and $3380^{+50}_{-60}$ K. The uncertainties represent the 68\% posterior confidence interval, which only includes formal uncertainties. The fit returns radii of $1.48^{+0.10}_{-0.09}$ \rsun\ and $1.42\pm0.09$ \rsun\ for the primary and secondary, respectively. With these posteriors we derive bolometric luminosities of $0.25\pm0.02$ and $0.23\pm0.02$ $L_\odot$ for the primary and secondary, respectively, using the Stefan-Boltzmann law. From this analysis we adopt the stellar radii and luminosities and present them in Table \ref{tab:char}, but adopt the \teff\ value derived in Section \ref{model_spec}.

The uncertainties of this fit are likely underestimated. The spectroscopically informed \teff\ prior constrains the posterior, which would favor lower temperatures in its absence. The sources of systematic uncertainties described in Section \ref{model_spec} are also at play here. A conservative $\pm$100 K uncertainty on the \teff\ is more appropriate, placing this value in fair agreement with the spectroscopically determined value. This value is placed in parentheses in Table \ref{tab:char}. We estimate the associated systematic uncertainty on the radius and luminosity values by smoothing the \teff-$R$ and \teff-$L$ posteriors with a Gaussian kernel. We set the kernel size to achieve a $\pm$100 K \teff\ uncertainty and adopt the 68\% confidence intervals along the $R$ and $L$ axes in order to capture the covariance between the parameters. These uncertainties are also included in the parentheticals in Table \ref{tab:char}. 

We map our SED fit \teff\ and $L$ distributions in the HR diagram to stellar mass, interpolating isochrones from the Dartmouth Stellar Evolution Program \citep[DSEP;][]{Dotteretal2008} using the {\tt scipy griddata} function. They correspond to a mass ratio of $1.0\pm0.1$ and a total mass of $0.58\pm0.05$ $M_{\odot}$, taking the median and standard deviation of the distributions. We confirm that the output component ages are coeval within observed ranges ($\sigma_{\Delta|log_{10} \tau|} < 0.16$ dex; \citealt{Kraus&Hillenbrand2009}), and find an age of $1.0\pm0.3$ Myr, which is in good agreement with the color-magnitude isochronal age of FO Tau's Taurus kinematic subgroup \citep[$1.34^{+0.18}_{-0.19}$ Myr;][]{Krolikowskietal2021}.

To estimate model-dependent bias in these measurements, we compare the mass ratio and total mass values to those derived from different model suites, and also assess their agreement with dynamical mass measurements from the literature. For the mass ratio, we find consistent values across seven other stellar evolution models (MESA Isochrones and Stellar Tracks \citep[MIST;][]{Dotteretal2016,Choietal2016}, BHAC 2015 \citep{Baraffeetal2015}, DSEP-magnetic \citep{Feiden&Chaboyer2012,Feiden&Chaboyer2013,Feiden2016}, SPOTS ($f_{spot} = $ 0, 0.17, 0.34, 0.51) \citep{Somersetal2020}). As such, we adopt the value above as the mass-ratio prior in our orbit fit. 

For the total mass, we find that values are highly model dependent. The standard stellar evolution models are largely self consistent (DSEP, MIST, BHAC, SPOTS $f_{spot}=0$), while the models that include prescriptions for the effect of magnetic fields (DSEP-magnetic, SPOTS $f_{spot} >$ 0) predict masses up to 65\% higher. Empirically, model agreement with dynamical masses is mass-dependent at young ages. For solar-type stars (FGK spectral types), the DSEP-magnetic models typically perform better than standard models \citep{Simonetal2019,Davidetal2019a,Braunetal2021}. At lower masses, mostly as the outcome of small number statistics, it is unclear which model predictions to favor. Although the results are on the border of statistical significance, the studies cited above suggest that the magnetic models overestimate the masses of objects $<0.4$ \msun\ based on their HR diagram locations. \citet{Rizzutoetal2016,Rizzutoetal2020a}, for instance, find that even the standard DSEP models over-predict dynamical measurements of binary total masses by 25\% to 85\% below 1 \msun. This does not suggest that the standard models provide a more physically realistic description of low-mass stars (which are certainly magnetic), only that there appears to be no single set of models that performs best at all masses and ages. In the case of young, low-mass stars, the short comings of standard models may simply conspire to provide more accurate mass predictions from the HR diagram than magnetic models. 

Given the variation between models, and evidence for systematic offsets with measured dynamical masses, we adopt the total-mass mean from the DSEP model above and a 1$\sigma$ width of 20\% (i.e., $\mathcal{N}(0.58,0.12)$) as our prior in the orbit fit. With this prior, we provide support for a scenario in which models with magnetic prescriptions are more accurate and one in which some or all models over-predict the masses of young, low mass stars. 

\begin{deluxetable*}{l c c c}
\tablecaption{Known and Derived Properties of FO Tau
\label{tab:char}}
\tablewidth{0pt}
\tabletypesize{\footnotesize}
\tablecolumns{4}
\phd
\tablehead{
  \colhead{Parameter} &
  \colhead{FO Tau A} &
  \colhead{FO Tau B} &
  \colhead{Reference}
}
\startdata
\multicolumn{4}{l}{\textbf{Identifiers}} \\
\gaia\ DR3& \multicolumn{2}{c}{163183644576299264}& \gaia\ DR3 \\
2MASS     & \multicolumn{2}{c}{04144928+2812305}  & 2MASS \\
TIC       & \multicolumn{2}{c}{56627416}          & \citet{Stassunetal2018} \\
\hline
\multicolumn{4}{l}{\textbf{Astrometry, Distance, \& Photometry}} \\
$\alpha$ RA (J2000)          & \multicolumn{2}{c}{04:14:49.3} & \gaia\ DR3\\
$\delta$ Dec (J2000)         & \multicolumn{2}{c}{+28:12:30.5} & \gaia\ DR3\\
$\mu_\alpha$ (mas yr$^{-1}$) & \multicolumn{2}{c}{$6.8\pm0.3$} & \gaia\ DR3\\
$\mu_\delta$ (mas yr$^{-1}$) & \multicolumn{2}{c}{$-23.9\pm0.2$} & \gaia\ DR3\\
$\varpi$ (mas)               & \multicolumn{2}{c}{$7.3\pm0.2$} & \gaia\ DR3\\
$RUWE$                       & \multicolumn{2}{c}{11.499} & \gaia\ DR3\\
Distance (pc)                & \multicolumn{2}{c}{$135^{+4}_{-3}$} & \citet{Bailer-Jonesetal2021} \\
$F_{\rm phot}$ at 6100 \AA\ (10$^{-14}$ \funit) & $0.66\pm0.11$ & $0.62\pm0.11$ & \citet{HartiganKenyon2003} \\
$F_{\rm phot}$ at 7510 \AA\ (10$^{-14}$ \funit) & \multicolumn{2}{c}{$6.9\pm1.2$} & \citet{HerczegHillenbrand2014} \\
$F_{\rm phot}$ in $H$-Band (10$^{-14}$ \funit) & $2.3\pm0.3$ & $2.2\pm0.3$ & This Work \\
$\Delta J$ (mag) & \multicolumn{2}{c}{$0.077\pm0.014$} & This Work \\
\hline
\multicolumn{4}{l}{\textbf{Stellar Properties}} \\
\teff\ (K)$^a$            & $3475\pm50(\pm100)$ & $3450\pm50(\pm100)$ & This Work\\
$R_\star$ ($R_\odot$)$^b$     & $1.48^{+0.10}_{-0.09} (\pm0.19)$ & $1.42\pm0.09 (\pm0.18)$ & This Work\\
$L_\star$ ($L_\odot$)$^b$     & $0.25\pm0.02 (\pm0.05)$ & $0.24\pm0.02 (\pm 0.05)$ & This Work\\
$M_\star$ ($M_\odot$)               & ${0.35}_{-0.05}^{+0.06}$ & ${0.34}\pm{0.05}$ & This Work\\
\logg                               & $3.9\pm0.1$ & $3.8\pm0.1$ & This Work\\
B (kG)                              & $1.8\pm0.1$ & $1.8\pm0.1$ & This Work\\
\vsini\ (\kms)                      & $14.0\pm1.0$ & $13.5\pm0.8$ & This Work\\
Veiling at 15600 \AA                & $0.35\pm0.05$ & $0.27\pm0.05$ & This Work \\
\mdot$_\star$ ($M_\odot$ yr$^{-1}$) & $2.6\times10^{-8}$ & $1.7\times10^{-8}$ & \citet{HartiganKenyon2003} \\
Spectral Type (NIR)                 & M 2.5 & M 2.5 & This Work \\
Spectral Type (optical)             & M 3.5 & M 3.5 & \cite{HartiganKenyon2003} \\
\hline
\multicolumn{4}{l}{\textbf{Binary Orbital Properties}} \\
$T_0$ (JD)                          & \multicolumn{2}{c}{${2465000}_{-3000}^{+2000}$} & This Work \\
$P$ (yr)                            & \multicolumn{2}{c}{${120}\pm20$}    & This Work \\
$a$ (AU)                            & \multicolumn{2}{c}{${22}_{-1}^{+2}$} & This Work \\
$e$                                 & \multicolumn{2}{c}{${0.21}_{-0.03}^{+0.04}$}      & This Work \\
$i_{\rm orbit}$ ($^\circ$)          & \multicolumn{2}{c}{${33}_{-5}^{+4}$}       & This Work \\
$\Omega$ ($^\circ$)                 & \multicolumn{2}{c}{${303}_{-10}^{+7}$}      & This Work \\
$\omega_A$ ($^\circ$)               & \multicolumn{2}{c}{${30}_{-20}^{+30}$}              & This Work \\
Minimum Separation ($a$($1-e$); AU) & \multicolumn{2}{c}{${18}\pm+1$}               & This Work \\
\hline
\multicolumn{4}{l}{\textbf{Disk Properties}} \\
1.3mm $F$ (mJy)                  & $2.96\pm0.07$ & $2.69\pm0.07$ & This Work \\
1.3mm Peak $I$ (mJy beam$^{-1}$) & $1.31\pm0.02$ & $1.28\pm0.02$ & This Work \\
log$_{10}(M_{\rm dust}/M_\odot)$ & $\gtrsim -5.26$ & $\gtrsim -5.29$ & This Work\\
$i_{\rm disk,\ dust}$ ($^\circ$)        & $27.3 \pm 0.5$      & $26 \pm 1$   & This Work \\
PA$_{\rm disk,\ dust}$ ($^\circ$)       & $121 \pm 1$   & $121 \pm 2$ & This Work \\
$R_{\rm eff,\ 95\% \ dust}$ (AU)             & $3.7 \pm 0.1$ & $3.6 \pm 0.5$ & This Work \\
$i_{\rm disk,\ CO}$ ($^\circ$)        & $27^{+7}_{-6}$      & $40^{+30}_{-20}$   & This Work \\
PA$_{\rm disk,\ CO}$ ($^\circ$)       & $120^{+30}_{-20}$   & $120^{+100}_{-50}$ & This Work \\
$R_{c,\ {\rm CO}}$ (AU)                   & $10 \pm 2$ & $8^{+4}_{-3} $ &This Work \\
$\Omega_{\rm disk}$ ($^\circ$)   & $300^{+30}_{-20}$   & $300^{+100}_{-50}$ & This Work \\
$v_{\rm sys}$ (\kms\ LSRK)       & $8.2^{+0.6}_{-0.5}$ & $6 \pm 2$            & This Work \\
\enddata
\tablecomments{($a$) Spectroscopically determined; additional uncertainty from systematic and astrophysical effects are included in the parenthetical (see Section \ref{model_spec}). ($b$) Determined from SED fitting in Section \ref{model_flux}.}
\end{deluxetable*}

\subsection{Relative Stellar Radial Velocity}
\label{spec_rv}

With angularly-resolved NIRSPEC spectra, we can compute the relative RV of the FO Tau A and B components, which provides a valuable constraint when fitting the orbital solution. We measure the relative RV by computing the spectral-line broadening function \citep[BF;][]{Rucinski1992} between the primary and secondary spectra. An in-depth description of the BF, as implemented here, can be found in \citet{Tofflemireetal2019}. In short, the BF is the linear inversion between a target and template spectrum and is typically used with an observed spectrum and a narrow-line, synthetic template. In that case, the BF represents a reconstruction of the average RV profile of photospheric absorption lines, providing the means to measure the star's RV and \vsini. In the present case, where the primary and secondary spectra are very similar, the BF of these two spectra returns a Gaussian profile centered at the RV offset between the two spectra with a width of the spectral resolution. 

We compute the BF for four, $\sim$100 \AA\ regions using the {\tt saphires} python package \citep{Tofflemireetal2019}. The regions are devoid of telluric contamination and together span 15450 to 16280 \AA. A Gaussian profile is fit to each region and we take the mean and standard deviation of these measurements as our relative RV and its internal RV uncertainty: $RV_B-RV_A = -1.8\pm0.2$ \kms. The negative value indicates that FO Tau B is moving toward Earth compared to FO Tau A.

\subsection{Binary Orbital Solution}
\label{orbit}

We fit the FO Tau orbital solution using the {\tt orvara} python package \citep{Brandtetal2021}. The fit uses a parallel-tempering MCMC framework \citep{Foreman-Mackeyetal2013,Vousdenetal2016} with 10 temperatures, 100 walkers, each taking 5$\times$10$^5$ steps. The orbit model includes nine parameters: the primary mass ($M_A$), secondary mass ($M_B$), semi-major axis ($a$), eccentricity ($e$), primary star's argument of periastron ($\omega_A$), inclination ($i$), position angle of the ascending node ($\Omega$), the reference longitude at 2010.0 ($\lambda_{ref}$), and the system parallax ($\varpi$). The eccentricity and argument of periastron are fit as $\sqrt{e}$ sin $\omega_A$ and $\sqrt{e}$ cos $\omega_A$. For reference, the $\Omega$ convention used here is the angle measured east from north where the secondary crosses the plane of the sky toward the observer. Gaussian priors are placed on the parallax (\gaia\ EDR3), and on the mass ratio ($q$) and total mass ($M_{\rm total}$) hyper-parameters, following the modeling in Section \ref{model_flux}. The chains for each parameter are saved every 50th step and chain convergence is assessed following the method outlined in Section \ref{model_flux}.

We include the relative RV measurement from Section \ref{spec_rv} in our fit, but inflate its uncertainty to include astrophysical RV jitter. Both stellar components are actively accreting and likely spotted, resulting in surface heterogeneities that drive RV scatter. With only one RV epoch we cannot measure a jitter value for either star, nor can we fit for it within the orbital solution. As such, we adopt an empirical NIR RV jitter value from the \citet{Crockettetal2012} RV survey of T Tauri stars. The median NIR jitter of their nine-star sample is $\sim 400$ m s$^{-1}$, which we add in quadrature to our relative RV error.

\begin{figure}[t!]
\begin{center}
\includegraphics[width=0.49\textwidth]{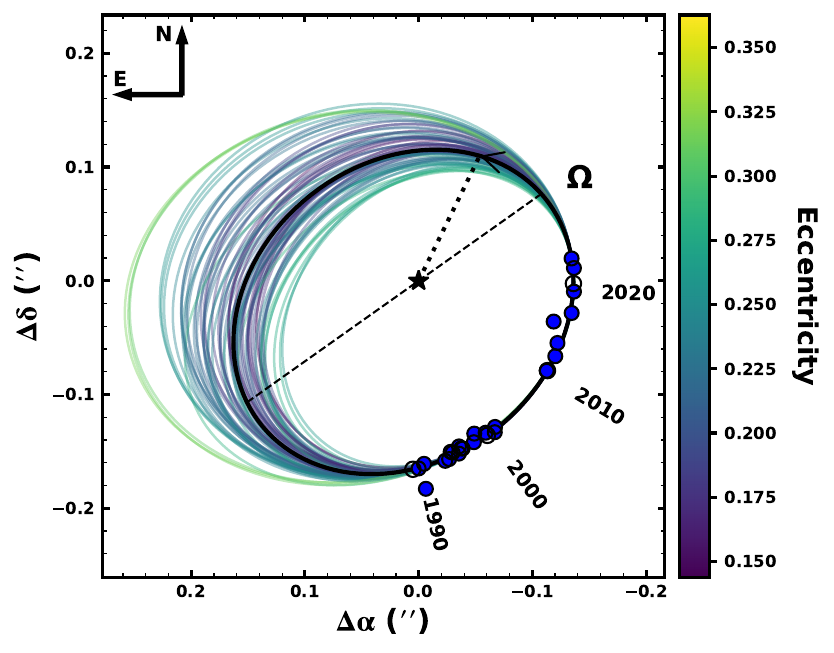}
\caption{On-sky projection of the FO Tau orbit. Observations are shown in blue circles. The orbit with the highest likelihood is shown in black. The dotted line signifies periastron passage. The dashed line is the line of nodes, with the $\Omega$ symbol signifying the ascending node. Random draws from the posterior distribution are shown in the background, color coded by the orbital eccentricity. The secondary is furthest from Earth in the South-West portion of the orbit and closest in the North-East.}
\label{fig:orbit}
\end{center}
\end{figure}

\begin{figure}[h!]
\begin{center}
\includegraphics[width=0.49\textwidth]{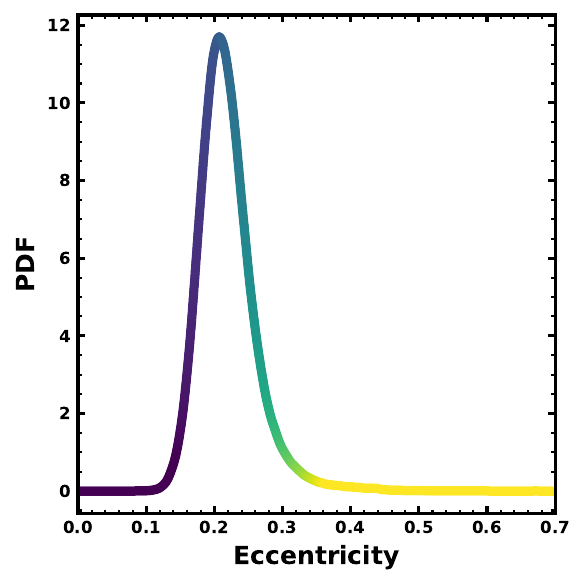}
\caption{FO Tau orbital eccentricity posterior. The probability distribution function is represented with a Gaussian kernel-density estimate. Line color matches the color bars in Figures \ref{fig:orbit} and \ref{fig:orvara}.}
\label{fig:ecc}
\end{center}
\end{figure}

\begin{figure}[h!]
\begin{center}
\includegraphics[height=0.835\textheight]{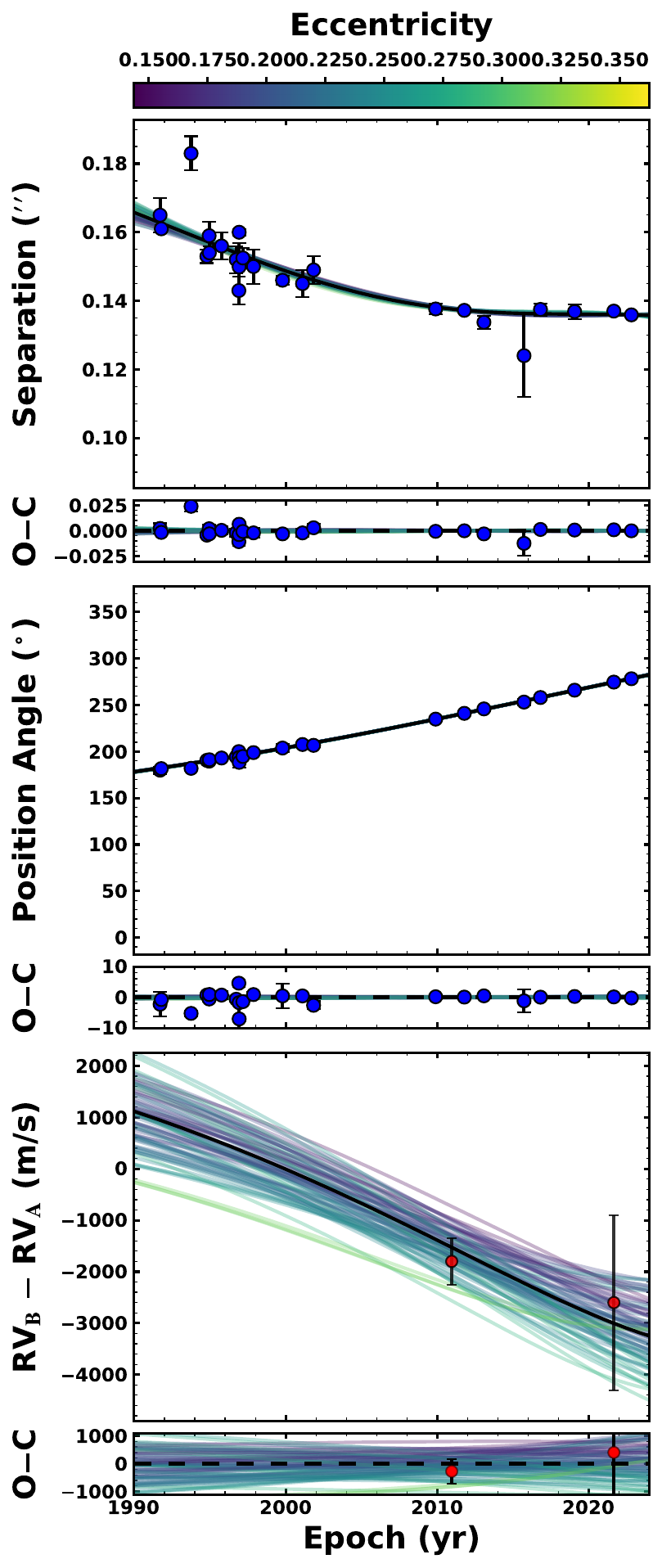}
\caption{The FO Tau orbit fit. As a function of time, we present the projected separation, position angle, and radial-velocity difference from top to bottom, each with its associated residuals. Astrometry measurements are show as blue circles. Radial-velocity difference measurements are shows as red circles. The highest-likelihood orbital solution is shown in black. Random posterior draws are shown in the background, color coded by the orbital eccentricity.}
\label{fig:orvara}
\end{center}
\end{figure}

We perform an initial orbit fit using the relative astrometry data in Table \ref{tab:AO} and the NIRSPEC relative RV. This fit provides posteriors for the stellar masses, which inform our modeling of the ALMA CO visibilities in Section \ref{disk_fit}. The disk model, in turn, fits a center of mass velocity for each component, providing an additional relative RV epoch. Our final fit includes the ALMA relative RV, which breaks degeneracies in the radial component of the orbital motion. Figure \ref{fig:orbit} presents the on-sky orbital solution. Our relative astrometry is shown in blue circles with the highest likelihood orbit in black. Dashed and dotted lines mark the line of nodes and argument of periastron, respectively. The $\Omega$ symbol marks the ascending node. Colored lines represent random draws from the parameter posteriors, color-coded by their eccentricity. FO Tau B is currently on the far side of its orbit (i.e., further from Earth than FO Tau A). The derived eccentricity posterior, following the same color scheme, is provided in Figure \ref{fig:ecc}. Figure \ref{fig:orvara} presents our full measurement set as a function of time compared to the orbit fit. Select parameters of the final fit are included in Table \ref{tab:char}. A full table of the fit parameters and their priors (Table \ref{tabap:orbit}), as well as a corner plot (Figure \ref{fig:orbit_corner}) are provided in Appendix \ref{ap:orbit}. 

We note that near-term improvement in the orbital solution would benefit most from additional relative-RV epochs, where the model is less constrained. Spatially-resolved, high-resolution spectra from AO facilities are required to make these measurements.

\

\subsection{Photometric Variability}

\begin{figure}[t!]
\begin{center}
\includegraphics[width=0.48\textwidth]{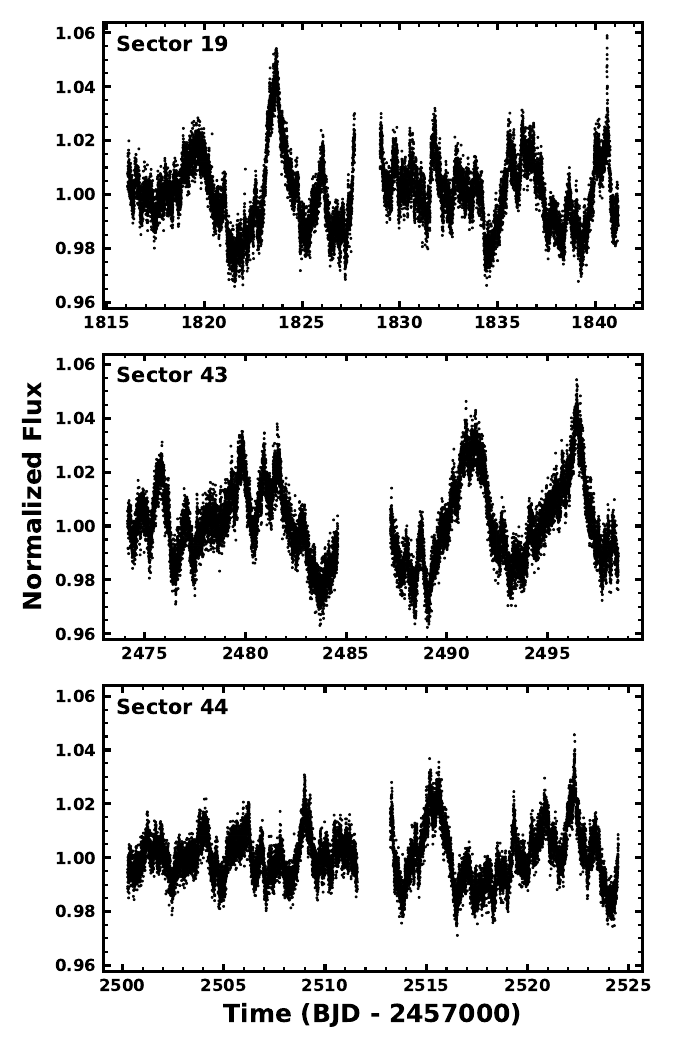}
\caption{\tess\ PDCSAP light curves of FO Tau. The variability is dominated by stochastic bursts associated with accretion and inner disk related processes.}
\label{fig:lc}
\end{center}
\end{figure}

Figure \ref{fig:lc} presents spatially unresolved FO Tau light curves from three \tess\ sectors. The variability is dominated by quasi-periodic brightening events that last multiple days. \citet{Robinsonetal2022} define the system as a ``burster'', based on the Sector 19 light curve using the $Q$ (periodicity) and $M$ (symmetry) variability metrics developed by \citet{Codyetal2014}. The burster population is characterized by high accretion rates (large UV continuum excesses) and warm inner disks (NIR excess). Variability in the \tess\ bandpass for these objects does not directly trace the accretion rate as measured at shorter wavelengths \citep{Robinsonetal2022}, and is likely the contribution of accretion and inner disk related processes. Their periodogram analysis produced a peak at $\sim4.2$ d, in agreement with analysis of TReS light curves in \citet{Xiaoetal2012}. We analyze the \tess\ light curves here in the search for periodic signals that could be plausibly linked to the stellar rotation signal. 

We use the Lomb-Scargle periodogram \citep[LSP;][]{Scargle1982} and auto-correlation function (ACF) to search for periodic behavior in the full three-sector light curve (19, 43, and 44) and in multiple smaller regions, specifically, the individual sectors, the union of back-to-back Sectors 43 and 44, and a subset of Sectors 43 and 44 that excludes the two large events at the end of Sector 43. We repeat the analysis for both the SAP and PDCSAP reductions given that additional systematics removed in the latter have been shown to affect rotation period measurements for stars with periods between 6--12 d (e.g., returning $P_{\rm rot}/2$; \citealt{Newtonetal2022,Magauddaetal2022}). 

We do not find a single period across the temporal regions analyzed that has a consistently dominant LSP peak with positive auto-correlation, regardless of the reduction. The $\sim$4.2 d peak, for instance, is only present in the Sector 19 PDCSAP light curve. The strongest LSP peak is found in Sector 43 at 5.2 d, which is largely driven by the two brightening events in the latter half of the sector (Figure \ref{fig:lc}, middle panel). Periodogram peaks near $\sim5.8$ d are present in multiple time ranges (Sector 19, 44, 43+44, and the 43+44 subset), but are not consistently the primary peak. Analyzing the full light curve, we find the strongest LSP peak at 5.9 d and a single ACF peak at 5.0 d. 

The present analysis is suggestive of some astrophysical periodicity between 4 and 6 d; however, we cannot confidently assign it an origin. Assuming some component of the periodicity is stellar rotation, for the stars' \vsini\ and radius values, this period range would correspond to inclinations between $\sim50^\circ$ to $75^\circ$. Although the formal uncertainty on this derivation is large \citep[e.g.,][]{Masuda&Winn2020}, it would disfavor alignment with the binary orbit. Conversely, if we assume the stars inherit the binary orbital inclination, the rotation period would be $\sim2.9$ d, which corresponds to negative ACF values (anti-correlation). Given the large amplitude variability and lack of spatial information, it seems most likely that the true stellar rotation periods are unrecoverable from the \tess\ light curves. As such we are not able to comment on the alignment of the stellar angular momenta with that of the binary orbit or the protoplanetary disks.

\section{Disk Properties}

\begin{figure}[t!]
\begin{center}
\includegraphics[width=0.48\textwidth]{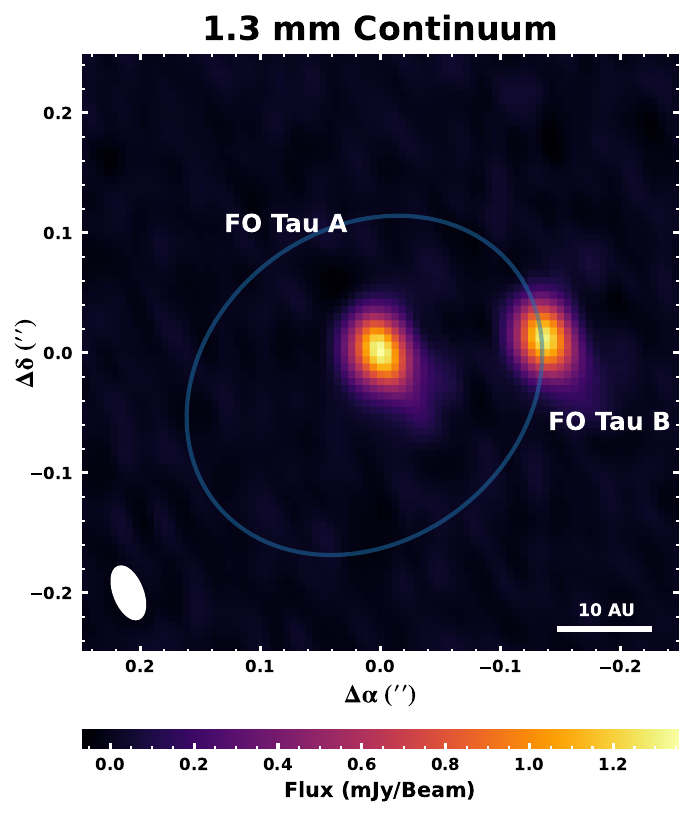}
\caption{The FO Tau 1.3 mm continuum map. Circumstellar dust is detected, but not clearly resolved, toward each of the FO Tau binary components. The highest-likelihood orbital solution is over-plotted in blue. The synthesized beam is presented in the bottom left corner and a 10 AU scale bar is shown in the bottom right.}
\label{fig:cont_map}
\end{center}
\end{figure}

\subsection{1.3 mm Continuum}
\label{continuum}

The 1.3 mm continuum image in Figure \ref{fig:cont_map} displays the clear detection of circumstellar dust associated with each stellar component. The dust emission is very compact and the observation is likely just on the boundary of resolving the dust disks. We fit the continuum map with a two-dimensional Gaussian profile for each component using the {\tt CASA} {\tt imfit} function. The fit provides the peak and integrated fluxes, which are presented in Table \ref{tab:char}. The centers of each component fit ($\alpha_{\rm A}$ = 04:14:49.30089, $\delta_{\rm A}$ = +28:12:29.9844; $\alpha_{\rm B}$ = 04:14:49.29056, $\delta_{\rm B}$ = +28:12:29.9957) are used to compute a separation and position angle for the fit to the binary orbit (Table \ref{tab:AO}; Section \ref{orbit}), and to fix the disk centers for our CO visibility modeling in Section \ref{disk_fit}. Depending on the initial guess provided to the function, the profiles are either narrowly resolved or consistent with point sources, indicating that we cannot robustly measure disk radii or their projections (e.g., inclination, position angle) in the image plane.

To measure continuum disk properties below the image-plane resolution, we forward model the continuum visibilities with an exponentially tapered power-law intensity profile,
\begin{equation}
    I(r) = I_0 \left(\frac{r}{R_c}\right)^{-\gamma_1} e^{-(r/R_c)^{\gamma_2}}.    
\end{equation}
Here, $\gamma_1$ is power-law index and $\gamma_2$ is the exponential-taper index at the transition radius, $R_c$. This model describes disks with sharply decreasing outer profiles which are seen in wider binary systems \citep{Manaraetal2019}, and allows us to make comparisons with studies using the same model \citep[e.g.,][]{Longetal2019,Manaraetal2019}. For each disk, this radial profile is modulated by an inclination, position angle, and positional offset resulting in an on-sky model image. Complex visibilities at the observed $uv$ baselines are computed using the {\tt galario} package \citep{Tazzarietal2018} and finally fit to our observations using {\tt emcee}. Our fit employs 80 walkers, each taking 7000 steps. The computational expense of this method does not lend itself to the chain auto-correlation convergence approach used in Section \ref{model_flux}, but inspection of the parameter chains shows clear convergence. The last 3500 steps are used for parameter estimation.  

The 68\% confidence interval of the parameter posteriors only includes systematic uncertainties and produces underestimated uncertainties. To obtain more robust errors, we fit each of the four spectral windows separately. Three are continuum spectral windows with 1.875 GHz of bandwidth, and the fourth is the CO-centered higher resolution spectral window with 0.9 GHz of bandwidth. Our adopted value is computed as the average of these four fits, weighted by their frequency bandwidth. The uncertainty is the bandwidth-weighted standard deviation.

For brevity, the detailed results of the fits are provided in Appendix \ref{ap:cont}. Table \ref{tabap:cont} contains the individual fits and the adopted values. Figure \ref{fig:cont_uv} presents the visibility and image-plane data-model comparisons, and displays the intensity profile for one of the spectral window fits. In short, the dust radii enclosed by 95\% of the total flux ($R_{\rm eff,\ 95\%}$) are $3.7\pm0.1$ pc and $3.6\pm0.5$ AU, the inclinations are $27.3\pm0.5^\circ$ and $26\pm1^\circ$, and positions angles are $121\pm1^\circ$ and $121\pm2^\circ$ for FO Tau A and B, respectively. These results are included in Table \ref{tab:char}.

The integrated 1.3 mm fluxes are consistent with those from \citet{Akesonetal2019} that were observed with lower angular resolution. Under the assumption that the dust emission is optically thin, we can compute a dust mass with the following equation:
\begin{equation}
    M_{\rm dust} = \frac{F_\nu d^2}{\kappa_\nu B_\nu(T_{\rm dust})},
\end{equation}
where $F_\nu$ is the 230 GHz flux density, $\kappa_\nu$ is the dust opacity (2.3 cm$^2$ g$^{-1}$ at 230 GHz; \citealt{Andrewsetal2013}) and $B_\nu(T_{\rm dust})$ is the Planck function evaluated at 230 GHz for the dust temperature, $T_{\rm dust}$. The dust temperature is computed assuming $T_{\rm dust} \sim 25(L_\star/L_\odot)^{1/4}$ K \citep[assumes optically thin dust]{Andrewsetal2013}. Including an additional 5\% uncertainty on the ALMA absolute flux calibration, we compute dust masses of $5.4(\pm0.4)\times10^{-6}$ \msun\ and $5.1(\pm0.4)\times10^{-6}$ \msun\ for the primary and secondary, respectively. These values are taken to be lower limits on the total dust mass considering how compact the disks are, and that the inner regions of protoplanetary disks have been shown to have non-negligible dust optical depths \citep[e.g.,][]{Huangetal2018,Birnstieletal2018}. We do not attempt to compute a total disk mass for this reason. 

We also note that at our detection limits, we do not detect any extended dust emission that might be indicative of a circumbinary disk. This is consistent with previous observations of FO Tau that were sensitive to larger spatial scales \citep{Akesonetal2019}.

\begin{figure*}[t!]
\begin{center}
\includegraphics[width=0.98\textwidth]{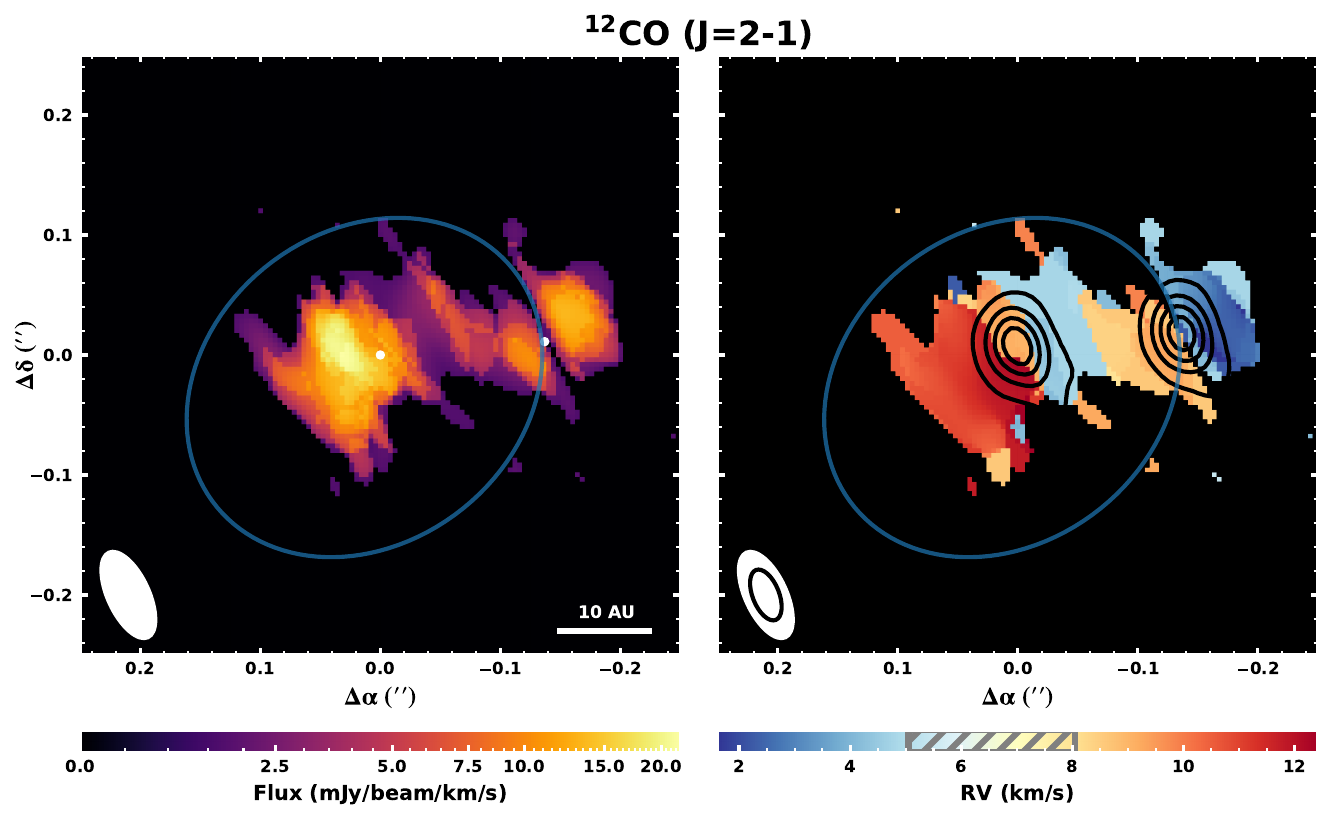}
\caption{Imaging of the FO Tau \co\ visibilities. {\bf Left:} The CO integrated intensity. All emission shown is $\geq$3$\sigma$ above the background noise. Absorption from intervening cloud material contaminates the north-west side of the primary disk and the south-east side and center of the secondary disk. The synthesized beam is presented in the bottom left and a 10 AU scale bar is provided in the bottom right. The highest-likelihood orbital solution is over-plotted in blue. White points mark the peak of continuum emission. {\bf Right:} The CO intensity weighted velocity map using the {\tt bettermoments} quadratic method \citep{Teague&Foreman-Mackey2018}. Extended CO gas disks are observed toward each component. Velocity channels heavily affected by cloud absorption are shown in the grey hatched region of the velocity color bar. Black contours are the 1.3 mm continuum emission, beginning at 5$\sigma$. The CO and continuum synthesized beams are presented in the white and black ellipses, respectively. In both panels a Keplerian mask, based on our best-fit disk model, is employed to reduce background noise for clarity (see Section \ref{co}).}
\end{center}
\label{fig:co_maps}
\end{figure*}

\subsection{$^{12}$CO J = 2--1}
\label{co}

Figure \ref{fig:co_maps} presents maps of \co\ emission. The left panel shows the CO integrated intensity, where extended circumstellar gas disks are observed toward each component. An asinh image stretch is used to emphasize faint emission. Cloud absorption is apparent in the map, affecting the north-west (blue shifted) side of the primary disk, and the south-east (red shifted) side and center of the secondary disk. The affected velocity channels correspond to $\sim$5 and 8 \kms\ (LSRK radio; hatched region in the right panel's velocity color bar). The right panel provides the intensity weighted velocity map  using the {\tt bettermoments} quadratic method \citep{Teague&Foreman-Mackey2018}. Our observations resolve the blue- and red-shifted orbital motion of each disk, but only marginally. The appearance of overlap of the two disks may be the result of the large beam size in comparison to their inherent size, or non-Keplerian flows that arise from tidal interactions. In both maps we employ a conservative Keplerian binary mask, based on the disk modeling in Section \ref{disk_fit}, to suppress background noise for clarity. No emission on the scale of the resolution element or larger is removed by the mask. An examination of CO emission at larger spatial scales did not reveal extended emission beyond the binary orbit that could be indicative of a circumbinary disk.

\subsection{Modeling the CO Gas Disk}
\label{disk_fit}

The spatial and spectral resolution of our CO data allows us to model the physical properties of the FO Tau protoplanetary disks. In the present work, we are primarily interested in the individual disk inclinations and their systemic velocities. At the same time, the sensitivity and angular resolution of our data limit our ability to constrain a full suite of disk parameters (or deviations from a single-star disk model). As such, we fit the continuum subtracted CO visibilities with an isolated disk model for each component, adopting the stellar parameters above and fixing or marginalizing over the more granular disk parameters. 

\begin{figure*}[th!]
\begin{center}
\includegraphics[width=0.98\textwidth]{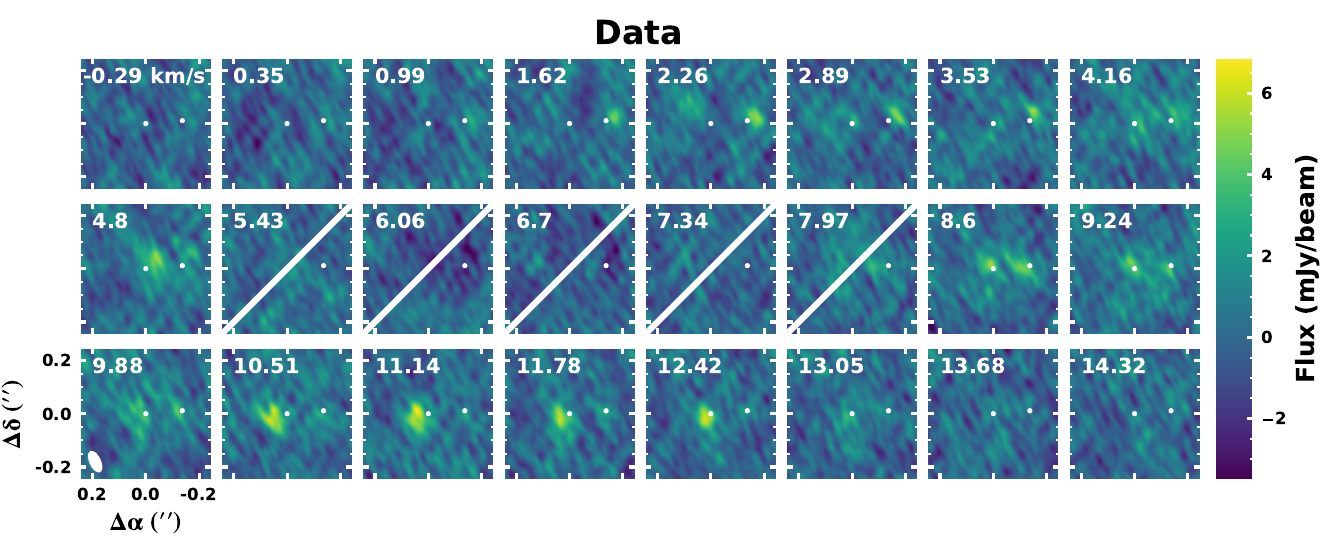}
\includegraphics[width=0.98\textwidth]{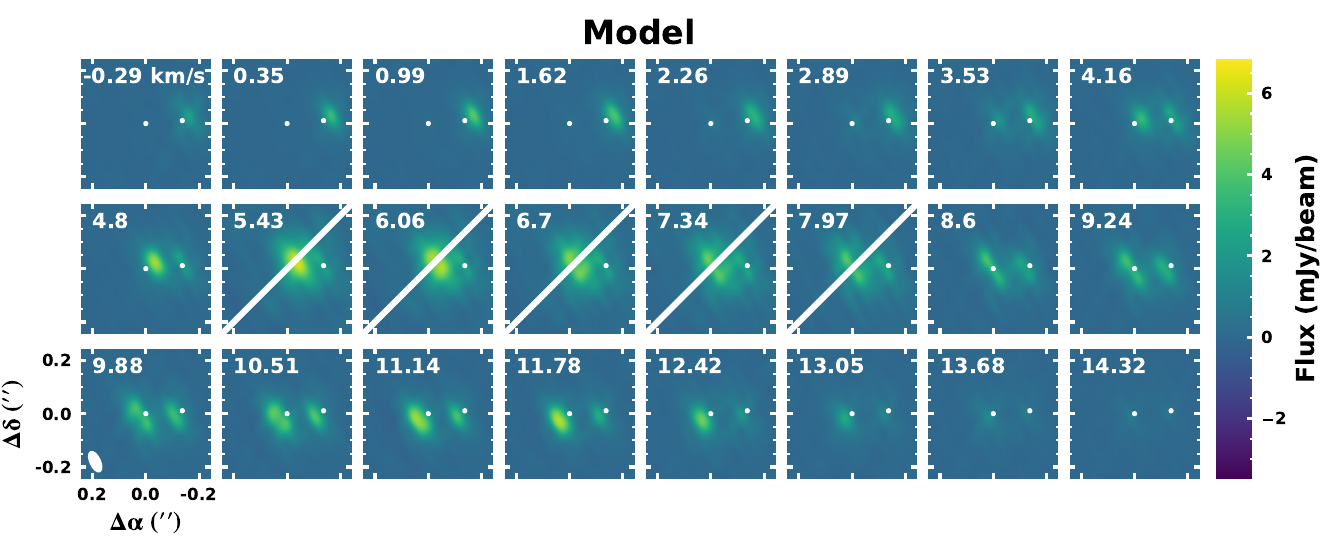}
\includegraphics[width=0.98\textwidth]{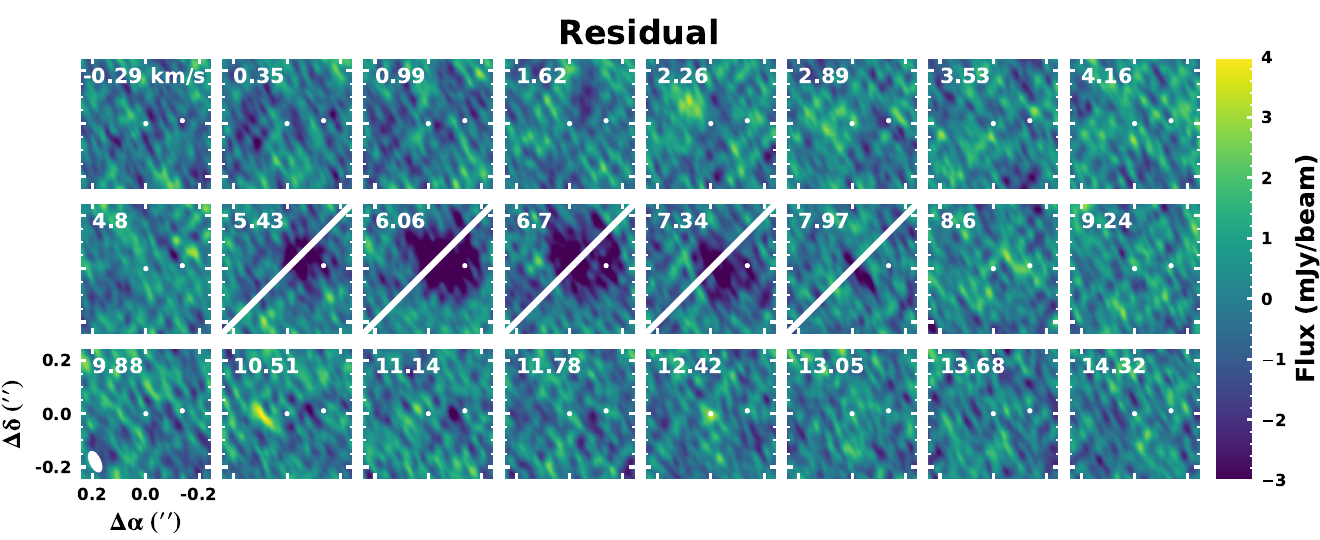}
\caption{\co\ velocity channel maps (LRSK radio) of the observations (top panel set), the best-fit model (middle panel set), and the data$-$model residuals (bottom panel set). Each panel provides the channel's central velocity in the top left corner. White points signify peaks in the continuum emission. Channels contaminated by cloud absorption have a diagonal white line.}
\label{fig:chan_map}
\end{center}
\end{figure*}

We adopt the parametric disk model outlined in  \citet{Flahertyetal2015}\footnote{\url{https://github.com/kevin-flaherty/disk_model3}}. Given our focus on the bulk disk properties, we provide a brief overview of the model here, but refer interested readers to the citation above or Appendix \ref{ap:disk} for more detail. The model assumes a temperature and gas-density structure in hydrostatic equilibrium \citep{Dartoisetal2003,Rosenfeldetal2013a}, and a Keplerian velocity field with modifications from gas pressure support. With the disk structure defined, line radiative transfer is computed along the line of sight assuming local thermodynamic equilibrium with Hanning smoothing applied to the resultant data cube. Each disk's structure is defined by twelve parameters, four of which are fit (but marginalized over), with the rest adopting fiducial values. The on-sky projection and radiative transfer of the disk are defined by nine parameters, six of which are defined by our ALMA observations or prior measurements, leaving three as free parameters: the disk inclination ($i_{\rm disk}$), its systemic velocity ($v_{\rm sys}$), and position angle (PA$_{\rm disk}$). For reference, the definition of PA in this model is 180$^\circ$ less than the disk's longitude of the ascending node, $\Omega_{\rm disk}$. A full list of the parameters, their adopted or fit values and any prior on their fit is included in Table \ref{tabap:disk} in Appendix \ref{ap:disk}.

The fit of the disk model is made in the $uv$ plane. We realize an on-sky model of two disks centered at the locations of the continuum centroids and compute complex visibilities at the observed $uv$ baselines and velocity channels using the {\tt galario} package. The model visibilities are fit to ALMA observations in a MCMC framework using {\tt emcee}. The fit is made to velocity channels between $-3.0$ and 17 \kms, ignoring channels between 5 and 8 \kms\ that have foreground cloud absorption. In total, 27 channels are used for the fit, each with a velocity width of 0.635 \kms. The fit is made with 80 walkers, each taking 7000 steps. The expense of the model computation does not allow us to assess convergence following the auto-correlation scheme in Section \ref{orbit}. Instead, we assess convergence by visually inspecting the parameter chains and confirming that parameter values are constant in 100 step increments moving backward from the last step. We conservatively use only the last 2000 steps to compute parameter values and 68\% confidence intervals. A subset of the fit parameters is provided in Table \ref{tab:char}, while the full suite is presented in Table \ref{tabap:disk} in Appendix \ref{ap:disk}, along with a corner plot.  

Channel maps of our CO observations, the best fit model, and fit residuals can be found in Figure \ref{fig:chan_map}. White circles indicate the location of the primary and secondary continuum peaks, and the four channels with white diagonal lines indicate the presence of cloud absorption. As seen in the residuals, the model is able to describe the bulk of the CO emission, however, there is extended emission, particularly in the 10.51 \kms\ channel that is not described by a Keplerian disk model. 

In the present fit, we marginalize over the disk radii given the angular resolution of our observations. We place a uniform prior between 1 and 14 AU in order to prevent the disks from overlapping along the line of sight. The median and 68\% confidence interval for the primary and secondary disk radii are $10 \pm 2$ and $8^{+4}_{-3}$ AU, respectively. These radii exceed the predicted dynamical truncation radii for FO Tau's orbital parameters (see Section \ref{disk_truncation}). In the event that there exist extended CO emission features (tidal or otherwise) that are not captured by our model, the disk radii and other parameters may be affected by our simple disk model attempting to fit extended emission. 

To explore this behavior further, we perform two additional MCMC fits that limit the upper bound of the radius priors to 8 and 5.5 AU. The disk parameters show some variation, but all have agreement within the 68\% confidence interval to the fiducial fit. The radius-constrained fits are unable to describe the emission at larger radii from the continuum peaks signaling that the large disk radii are not solely the result of the angular resolution of the observation.

\section{Discussion}
\label{discussion}

\subsection{The Binary-Disk Interaction}
\label{binarydiskinteraction}

In the following subsections we discuss our measurements of the FO Tau orbit and protoplanetary disks in the context of theory describing the binary-disk interaction. 

\subsubsection{Disk-Orbit Alignment}
\label{alignment}

\begin{figure*}[t!]
\begin{center}
\includegraphics[width=0.98\textwidth]{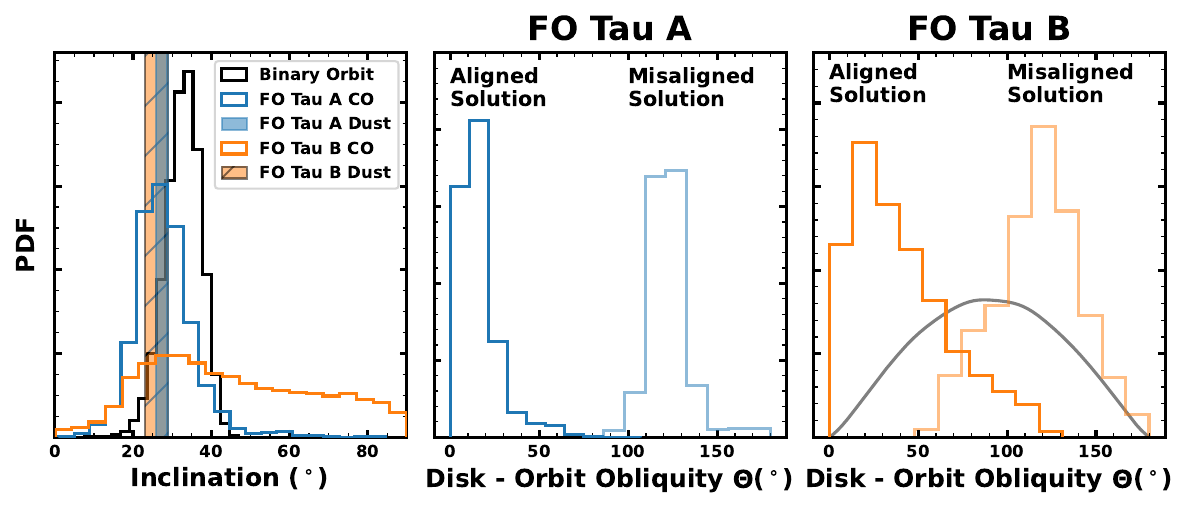}
\caption{{\bf Left:} Comparison of the projected inclination posteriors of the binary orbit and the individual protoplanetary disks. Measurements from both the continuum (dust) and CO visibility fits are included for each disk. {\bf Middle and Right:} Disk-orbit obliquity posteriors for the FO Tau A and FO Tau B CO disks, respectively. In each panel, the dark and faint lines represent the aligned and misaligned solutions, respectively (as labeled). A schematic displaying these scenarios in presented in Figure \ref{fig:schematic}. The grey curve in the right panel corresponds to the obliquity distribution of a randomly oriented disk.}
\label{fig:inc_posts}
\end{center}
\end{figure*}

The primary feature of the binary-disk interaction that the present study allows us to comment on is the alignment between the binary orbit and the individual circumstellar disks. We find evidence for alignment among the binary and protoplanetary orbital planes within the uncertainty of our measurements. The left panel of Figure \ref{fig:inc_posts} presents the inclination measurements for the three components. The orbit and CO disk inclinations are presented as the fit posteriors. The dust disk inclinations are presented as vertical bands with widths that correspond to 3$\sigma$. Our measurement for the secondary disk's CO inclination has the largest uncertainty, but the peak in the probability distribution aligns with that of the other more precise measurements. 

Agreement in the projected inclinations, however, does not necessarily correspond to alignment. A more direct measure of binary and disk orbital planes is their obliquity ($\Theta$), the angle between the binary and disk angular momentum vectors. Functionally:
\begin{equation}
\begin{split}
    {\rm cos}\ \Theta =\ &{\rm cos}\ i_{\rm disk}\ {\rm cos}\ i_{\rm orbit}\ + \\
    &{\rm sin}\ i_{\rm disk}\ {\rm sin}\ i_{\rm orbit}\ {\rm cos}(\Omega_{\rm disk} - \Omega_{\rm orbit}).
\end{split}
\end{equation}
The three-dimensional constraints on the binary orbit (direction of orbital motion, inclination, and longitude of the ascending node) allow for a direct measure of the binary orbit's angular momentum vector.
The disks, however, have a degeneracy in the direction of rotation. Counter clockwise rotation corresponds to an angular momentum vector pointing toward the observer. Clockwise rotation corresponds to an angular momentum vector pointing away the observer. Framed in another way, our observation does not reveal the near and far sides of the disks. This degeneracy can be lifted with future observations of disk gas at high angular resolution that resolve the front and back of the disk atmosphere \citep[e.g.,][]{Rosenfeldetal2013a}, or with angularly-resolved, scattered-light imaging that is sensitive to forward-scattered light \citep[e.g.][]{Avenhausetal2018}. 

\begin{figure*}[t!]
\begin{center}
\includegraphics[width=0.98\textwidth]{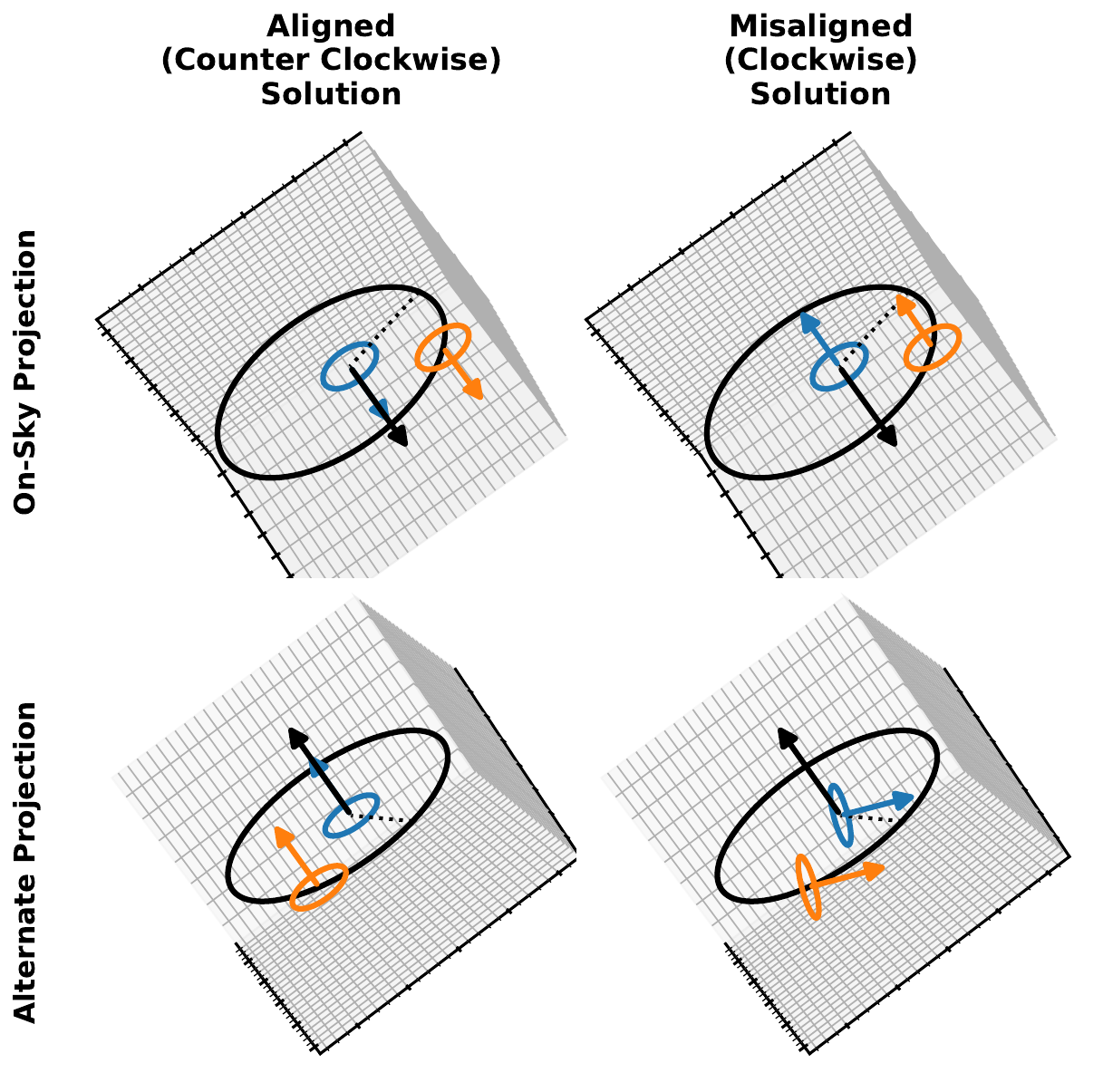}
\caption{Schematics of the binary-disk orbital planes for the two possible disk orientations allowed by our observation. The binary orbit is shown in black. The  primary and secondary disks in blue and orange, respectively. Arrows indicate the angular momentum vectors. The dotted line marks periastron passage. The left column is the aligned solution in which the disks are rotating counter clockwise with the binary orbit. The right column is the misaligned solution where the disks rotate clockwise, retrograde with the binary orbit.  The top row shows the on-sky projection where the projected disk inclinations are the same, despite having different directions of rotation. The bottom row presents an alternate projection where the difference between the solutions is more readily apparent.}
\label{fig:schematic}
\end{center}
\end{figure*}

Until then, two solutions exist: an aligned solution in which the disk rotates counter clockwise with the binary orbit, and a misaligned solution where the disk rotates clockwise, retrograde with the binary orbit. We compute the disk-orbit obliquity for both solutions and present them for FO Tau A and B in the middle and right panels of Figure \ref{fig:inc_posts}, respectively. The right panel also includes the distributions of obliquities for random disk orientations in grey, for reference. (For clarity, we only include the obliquity distributions for the CO disks.) A schematic aid to visualize these two scenarios is presented in Figure \ref{fig:schematic}. Although we cannot strictly rule out the pathological misaligned scenario, it seems unlikely given the consistency between projected inclinations and the longitudes of the ascending node for the orbit and disks.

Assuming the aligned solution, the immediate question becomes whether FO Tau formed in this way, or whether this quality is the result of subsequent dynamical evolution. If the former, it may be signaling a formation pathway that favors disk retention in close binaries. The fraction of young, close binaries with disks (or mature binaries with planets) might then reflect the frequency of this formation pathway. Coplanarity between the binary and disk planes, a small binary separation, a low eccentricity, and a stellar mass ratio near unity are all expectations from the disk-fragmentation paradigm \citep[e.g.,][]{Bate&Bonnell1997,Ochietal2005,Youngetal2015}. At the same time, stellar pairs that form at wider separations, in a core-fragmentation scenario, can rapidly migrate and achieve binary-disk alignment via dynamical interactions \citep{Zhao&Li2013,Leeetal2019,Guszejnovetal2023}. Additionally, disk fragmentation is expected to require massive disks \citep[e.g.,][]{Kratteretal2010,Zhuetal2012}, and for the relatively low mass in the FO Tau system, most Class 0/I protostellar analogs are compact and stable to gravitational collapse \citep{Tobinetal2020}. 

There exist many mechanisms in the literature that have been suggested to bring binary-disk systems into or out of alignment (e.g., envelope accretion, ejections, inclination oscillations, damping; see \citealt{Offneretal2022} and references therein). One mechanism we explore, given the small semi-major axis of the system, is the damping of mutual disk-orbit inclinations through viscous warped disk torques \citep{Bate2000,Lubow&Ogilvie2000}. Following \citet{Zanazzi&Lai2018}, for FO Tau's nominal parameters, the disk damping rate takes the form
\begin{equation}
\begin{split}
    \gamma_b = 4.5\times10^{-6}
    \left( \frac{\alpha}{10^{-2}} \right)
    \left( \frac{0.1}{h_{\rm out}} \right)^2
    \left( \frac{M_2}{0.34 \ M_\odot} \right)^2 \\
    \times
    \left( \frac{0.35 \ M_\odot}{M_1} \right)^{3/2}
    \left( \frac{22 \ {\rm AU}}{a} \right)^6
    \left( \frac{r_{\rm out}}{7 \ {\rm AU}} \right)^{9/2}
    {\rm yr}^{-1}
\end{split}
\end{equation}
where $\alpha$ is the viscosity parameter, $h_{\rm out}$ is the aspect ratio at the outer disk edge, $M_1$ and $M_2$ are the primary and secondary stellar masses, respectively, $a$ is the semi-major axis, and $r_{\rm out}$ is the outer disk edge. (Here, we have assumed the dimensionless viscous coefficient, $\mathcal{V}_{\rm b} \approx 1$; see Table~1 of \citealt{Zanazzi&Lai2018} for justification.) For the representative values above, the alignment timescale $(\gamma_b^{-1})$ is less than the age of the system ($\lesssim 1$ Myr), but would quickly exceed it for smaller viscosity values or larger disk aspect ratios. The latter may be particularly relevant in the light of numerical simulations by \citet{Picogna&Marzari2013} that find disks are dynamically heated by companions, especially in the vertical direction. 

The present analysis does not distinguish between formation or subsequent evolution as the source of FO Tau's disk-orbit alignment. A larger sample, particularly at larger separations (where the damping timescales are longer but orbital parameters are admittedly more uncertain), may help to address this question. FO Tau has the largest separation in our full ALMA sample, so our current data set is unlikely to address this ambiguity. It will, however, address whether disk-orbit alignment is common in protoplanetary disk-hosting binaries with small separations.

\subsubsection{Disk Truncation}
\label{disk_truncation}

Theory predicts that the binary orbit will drive resonances that truncate the outer edges of circumstellar disks \citep{Artymowicz&Lubow1994,Lubowetal2015,Miranda&Lai2015}. With some dependence on the binary orbital parameters, the disk parameters, and the mutual inclination between the two, the disk truncation radius is on the order of $\sim30\%$ of the orbital semi-major axis. \citet{Manaraetal2019} present an equation for the disk truncation radius following the work of \citet{Artymowicz&Lubow1994}, which has the form,
\begin{equation}
    R_{\rm trun} = \frac{0.49 a q^{-2/3}}{0.6q^{-2/3} + \ln(1+q^{-1/3})}\left( b e^c + 0.88\mu^{0.01}\right).
\end{equation}
Here, $a$ is the semi-major axis, $q$ is the stellar mass ratio ($M_B/M_A$), $e$ is the orbital eccentricity, and $\mu$ is the secondary to total-mass ratio ($M_B/(M_A+M_B)$). The remaining variables, $b$ and $c$, depend on $\mu$ and the disk Reynolds number. Using the range of $b$ and $c$ coefficient values compiled by \citet[their Appendix C.1]{Manaraetal2019}, we compute disk truncation radii of 5--6 AU for FO Tau's orbital parameters. 

The dust disk sizes we measure (95\% effective radius) are $\sim$4 AU. Radii below the dynamical prediction are consistent with the expectation that dust should be more compact than gas resulting from radial drift \citep[e.g.,][]{Ansdelletal2018}. In the case of close binaries, radial drift may be operating at higher rates given the impact of disk truncation \citep{Zagariaetal2021a,Rotaetal2022}.

As for the gas disk, our fit to the CO visibilities only loosely constrains the individual disk radii (Section \ref{disk_fit}), but favors values that are larger than those predicted above ($R_c\sim8-10$ AU). While the observations do resolve the CO emission, they do not have the angular resolution to distinguish whether the fit radii are indeed larger than truncation models predict, or whether our simple model is not capturing more complex emission. The latter seems likely given the extended emission described in the next sub-section, but determining the degree to which the radii may be inflated will require higher-angular-resolution observations. 

The average ratio of the gas to dust radius ($R_{CO}/R_{\rm dust}$) in wide binaries has been shown to be larger than single stars, when using the 95\% effective radius: 4.2 for binaries \citep{Rotaetal2022} and 2.8 for single stars \citet{Sanchisetal2021}. We do not measure the 95\% effective radius for the CO disks given the complicated on-sky projection, but from the critical radius of our CO visibility fit, we compute gas to dust disk radius ratios of $2.6\pm0.5$ and $2\pm1$ for FO Tau A and B, respectively. These values are below the population averages but are not outliers in the distribution of either binary or single stars.

Lastly, we note that mutual inclinations between the binary and disk planes can lead to larger disk truncation radii \citep{Miranda&Lai2015,Lubowetal2015}. However, the effect is relatively small; a $90^\circ$ mutual inclination would correspond to a $\sim30\%$ increase in the disk radius. Given that the best-fit disk radii are still larger than this prediction and our measurements are consistent with alignment, we do not find that the large gas-disk radii measured necessarily support this scenario.

\subsubsection{Extended CO Emission}

In the residual image of the CO visibilities fit (Figure \ref{fig:chan_map}, bottom panel set), we find an arc of extended emission to the east of the primary disk (LSRK $= 10.51$ \kms). The entire arc is more than $3\sigma$ above the background RMS noise, with a peak $>4\sigma$. The feature is also visible as a spur in the CO intensity and first moment maps (Figure \ref{fig:co_maps}).

The arc-like feature is reminiscent of the extended emission that has been seen in gas maps \citep{Rodriguezetal2018,Zapataetal2020,Cuelloetal2023} and scattered light imaging \citep{Zhangetal2023,Weberetal2023} of wide binary systems at similar ages to FO Tau. These observations can offer a lens into the broader, low-density environment and trace structures linked to formation, which are not probed by the larger dust grains with (sub-)mm continuum. Some of the systems referenced above may result from flybys \citep{Cuelloetal2023}, rather than more stable binary systems like FO Tau. Still, tidal features like extended spiral arms and bridges connecting circumstellar disks are also expected to arise from the binary-disk interaction \citep[e.g.,][]{Nelson2000,Zsometal2011,Mulleretal2012,Picogna&Marzari2013,Jordanetal2021}. The present feature, if resolved with future observations, may provide a test of the disk parameters (e.g., viscosity) that determine the size and locations of extended features.

\subsubsection{Future Sub-mm Observations}

Despite the challenges in observing compact binaries, FO Tau remains one of the best candidates to study the binary-disk interaction. Now with a full orbital solution, it is one of the widest binaries with precisely known orbital parameters and two bright circumstellar disks. Additionally, its relatively low orbital eccentricity, compared to the current sample of young binary orbits \citep{Prato2023}, should allow for larger disk truncation radii compared to other young systems. Future observations at higher angular resolution that target less optically-thick gas tracers will fully illuminate the discussion above.

\section{Summary \& Conclusions}

In this study we analyze the orbital, stellar, and protoplanetary disk properties of the young pre-main sequence binary, FO Tau. We have combined decades of high-angular-resolution imaging, angularly-resolved high-resolution NIR spectra, and (sub-)mm interferometry to develop a comprehensive view of the system's dynamical state. With this we can make some of the first tests of the binary-disk interaction and its effect on planet formation. 
The main conclusions of our work are as follows:

\begin{enumerate}
    \item FO Tau is a young binary system with a semi-major axis of 22 AU. The stars are near-equal mass, with dynamical masses of 0.35 \msun\ and 0.34 \msun for the primary and secondary, respectively. The orbital eccentricity (0.21) is low compared to other young binaries at similar separations.
    
    \item ALMA observations in Band 6 detect continuum and \co\ line emission, resolving the contribution from each binary component. The disks are compact and likely truncated by the binary orbit. Circumstellar dust disks are not resolved in the image plane ($\sim$5 AU resolution) and the CO gas disks are only marginally resolved. Foreground cloud absorption contaminates the CO emission from parts of both disks. The integrated 1.3mm fluxes for each component are near equal at $2.96\pm0.07$ and $2.69\pm0.07$ mJy for the primary and secondary, respectively.  

    \item Measuring inclinations of the gas and dust disks with fits made in the $uv$ plane, we find evidence for alignment between both protoplanetary disks and the binary orbit. Our gas-disk fit does not provide strong constraints on the disk radii, but suggests evidence for either larger radii than predicted by truncation models, or a more complex emission structure that is not captured by our model.
        
    \item We do not find evidence for a circumbinary disk or extended emission detected beyond the binary orbit in dust or CO above an RMS of 0.02 and 0.9 mJy/beam, respectively.
\end{enumerate}

FO Tau is the first young binary system where the properties of the circumstellar protoplanetary disks can be compared to precisely known orbital parameters. The synthesis of these results above points to the FO Tau system as a relatively placid environment (mutually aligned, low eccentricity). Most young binaries at this separation do not host disks, so it is tempting to tie this property to the retention of a substantial reservoir of circumstellar material, and, perhaps, a common quality of binaries that form and retain planets. Indeed, there is growing evidence for preferential alignment among circumstellar disks in wider binaries \citep[with unknown orbital parameters]{Jensenetal2020} and low mutual inclinations in the planet-binary orbital planes of field-age systems \citep{Dupuyetal2022,Christianetal2022,Behmardetal2022,Lesteretal2023}. Although only suggestive now, coupling ALMA interferometry with the growing number of young, disk-bearing binaries with known orbital parameters will facilitate the same detailed dynamical study presented here for a representative population. 

\facilities{ALMA, Texas Advanced Computing Center (TACC), Keck:II (NIRC2,NIRSPEC)}
\software{{\tt astropy} \citep{astropy1,astropy2}, 
{\tt emcee} \citep{Foreman-Mackeyetal2013}, 
{\tt saphires} \citep{Tofflemireetal2019},
{\tt scipy} \citep{scipy,scipy2},
{\tt RotBroadInt} \citep{Carvalho&Johns-Krull2023}.
}

\section*{acknowledgments}

We would like to thank the referee for a constructive and helpful report. BMT would like to thank Sam Factor, Kendall Sullivan, Neal Evans, and Stella Offner for helpful conversations. BMT is supported by the Heising-Simons Foundation's 51 Pegasi b Postdoctoral Fellowship in Planetary Astronomy. 

GHS acknowledges support from NASA Keck PI Data Awards administered by the NASA Exoplanet Science Institute (PI Schaefer; 2016B-N046N2, 2019B-N166). Time at the Keck Observatory was also granted through the NOIRLab (PropID: 2022B-970020; PI: G.\ Schaefer) supported by the NSF Mid-Scale Innovations Program. LP was supported in part by NSF awards AST-1313399 and AST-2109179. Contributions to the development of the synthetic spectral grid were made by Thomas Allen, Ian Avilez, Kyle Lindstrom, Cody Huls, and Shih-Yun Tang, all formerly at Lowell Observatory.

This paper makes use of the following ALMA data: ADS/JAO.ALMA\#2019.1.01739.S. ALMA is a partnership of ESO (representing its member states), NSF (USA) and NINS (Japan), together with NRC (Canada), MOST and ASIAA (Taiwan), and KASI (Republic of Korea), in cooperation with the Republic of Chile. The Joint ALMA Observatory is operated by ESO, AUI/NRAO and NAOJ. The National Radio Astronomy Observatory is a facility of the National Science Foundation operated under cooperative agreement by Associated Universities, Inc.

Some of the data presented herein were obtained at the W. M. Keck Observatory, which is operated as a scientific partnership among the California Institute of Technology, the University of California and the National Aeronautics and Space Administration. The Observatory was made possible by the generous financial support of the W. M. Keck Foundation. 

This research has made use of the Keck Observatory Archive (KOA), which is operated by the W. M. Keck Observatory and the NASA Exoplanet Science Institute (NExScI), under contract with the National Aeronautics and Space Administration.

The authors acknowledge the Texas Advanced Computing Center (TACC) at The University of Texas at Austin for providing HPC resources that have contributed to the research results reported within this paper\footnote{\url{http://www.tacc.utexas.edu}}.

Figures in this manuscript were created using color-impaired-friendly schemes from ColorBrewer 2.0\footnote{\url{https://colorbrewer2.org/}}.

The authors wish to recognize and acknowledge the very significant cultural role and reverence that the summit of Mauna Kea has always had within the indigenous Hawaiian community. We are most fortunate to have the opportunity to conduct observations from this mountain. 

We would like to acknowledge the Alabama-Coushatta, Caddo, Carrizo/Comecrudo, Coahuiltecan, Comanche, Kickapoo, Lipan Apache, Tonkawa and Ysleta Del Sur Pueblo, and all of the American Indian and Indigenous Peoples and communities who have been or have become a part of the lands and territories of Texas.


\appendix

\section{Binary Orbital Parameter Posteriors}
\label{ap:orbit}

\begin{figure*}[h!]
\begin{center}
\includegraphics[width=0.98\textwidth]{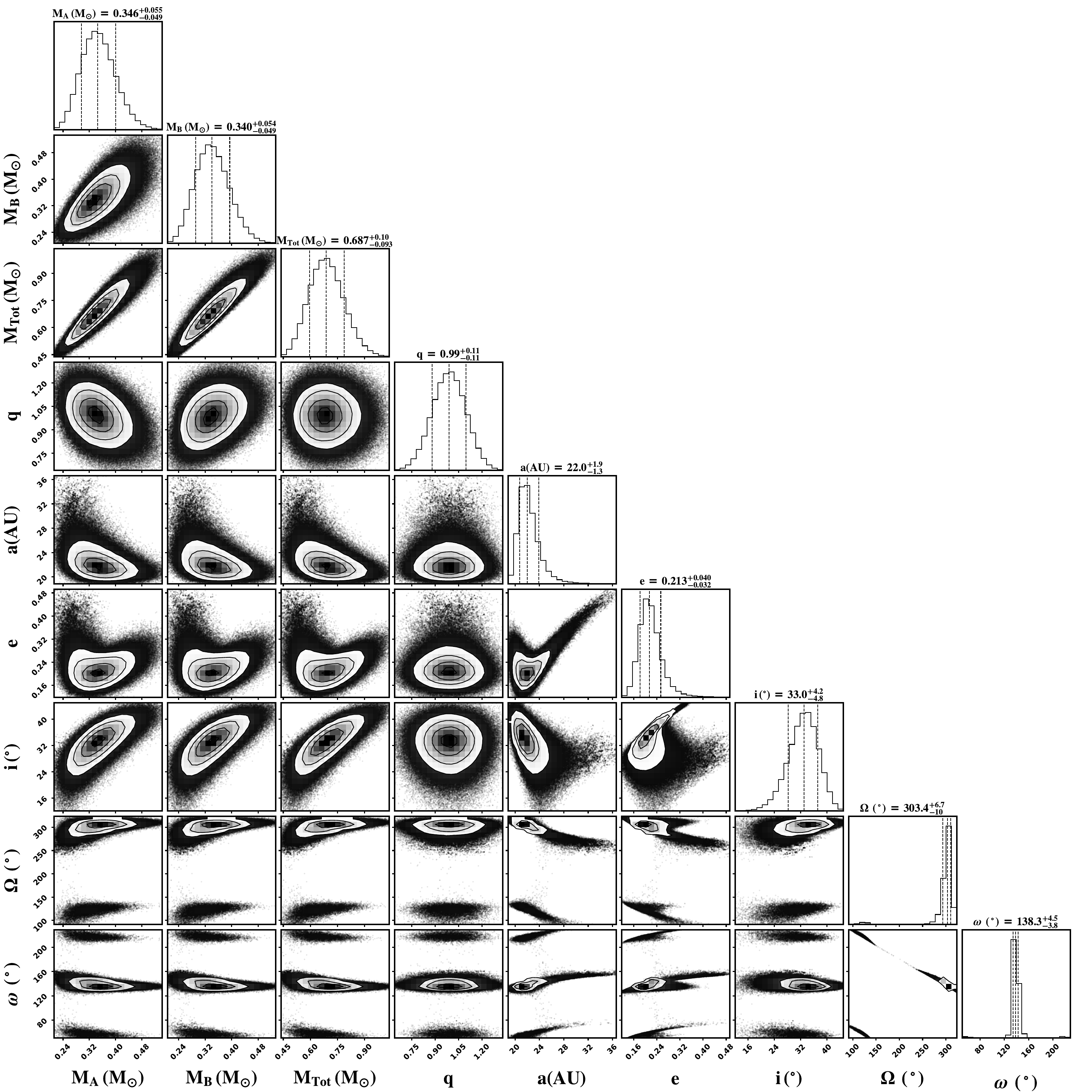}
\caption{Corner plot for parameters in the FO Tau orbit fit.}
\label{fig:orbit_corner}
\end{center}
\end{figure*}

This appendix section provides a summary of the parameters in the FO Tau orbit modeling and the full results of our fit. Table \ref{tabap:orbit} present the parameters that are fit, the hyper-parameters used to place additional priors on the fit, and the orbital parameters derived from the fit. A corner plot of the fit parameters is included in Figure \ref{fig:orbit_corner}.

\begin{deluxetable*}{l c c c}
\tablecaption{FO Tau Orbital Parameters
\label{tabap:orbit}}
\tablewidth{0pt}
\tabletypesize{\footnotesize}
\tablecolumns{4}
\phd
\tablehead{
  \colhead{Parameter} &
  \colhead{Prior} &
  \colhead{Value} &
  \colhead{Description} 
}
\startdata
\multicolumn{4}{l}{\textbf{Fit Parameters}} \\
$M_A \ (M_\odot)$               & Uniform & ${0.346}_{-0.049}^{+0.055}$ & Primary mass \\
$M_B \ (M_\odot)$               & Uniform & ${0.340}_{-0.049}^{+0.054}$ & Secondary mass \\
$a$ (AU)               & 1/$a$ (log flat) & ${22.0}_{-1.3}^{+1.9}$  & Semi-major axis \\
$\sqrt{e}$ sin $\omega_A$       & Uniform & ${0.19}_{-0.17}^{+0.15}$    & \\ 
$\sqrt{e}$ cos $\omega_A$       & Uniform & ${0.390}_{-0.082}^{+0.077}$ & \\
$i_{\rm orbit} \ (^\circ)$      &         & ${33.0}_{-4.8}^{+4.2}$  & Inclination \\
$\Omega \ (^\circ)$             & Uniform & ${303.4}_{-10}^{+6.7}$  & Position angle of ascending nodes \\
$\lambda_{ref} \ (^\circ)$      & Uniform & ${311.0}_{-8.7}^{+10}$  & Reference longitude at UTC 2010.0 \\
$\varpi$ (mas)   & $\mathcal{N}(7.3,0.2)$ & ${7.42}_{-0.20}^{+0.20}$    & Parallax \\
\hline
\multicolumn{4}{l}{\textbf{Hyper-Parameters}} \\
$M_{\rm total}=M_A + M_B \ (M_\odot)$ & $\mathcal{N}(0.57,0.11)$  & ${0.687}_{-0.093}^{+0.10}$ & Total Mass \\
$q=M_B/M_A$                           & $\mathcal{N}(1.00,0.05)$  & ${0.99}_{-0.11}^{+0.11}$ & Mass Ratio \\
sin $i_{\rm orbit}$                   & sini $i$, $i \in [0,180]$ & ${0.544}_{-0.072}^{+0.060}$ & \\
\hline
\multicolumn{4}{l}{\textbf{Derived Parameters}} \\
$P$ (yr) & & ${124}_{-15}^{+23}$ & Period \\
$T_0$ (JD) & & ${2464669}_{-2645}^{+1677}$ & Time of periastron passage \\
$e$ & & ${0.213}_{-0.032}^{+0.040}$ & Eccentricity \\
$\omega_A$ & & ${33}_{-22}^{+25}$ & Argument of periastron for primary orbit \\
\enddata
\end{deluxetable*}

\section{Continuum Disk Modeling}
\label{ap:cont}

\begin{figure*}[t!]
\begin{center}
\includegraphics[width=0.98\textwidth]{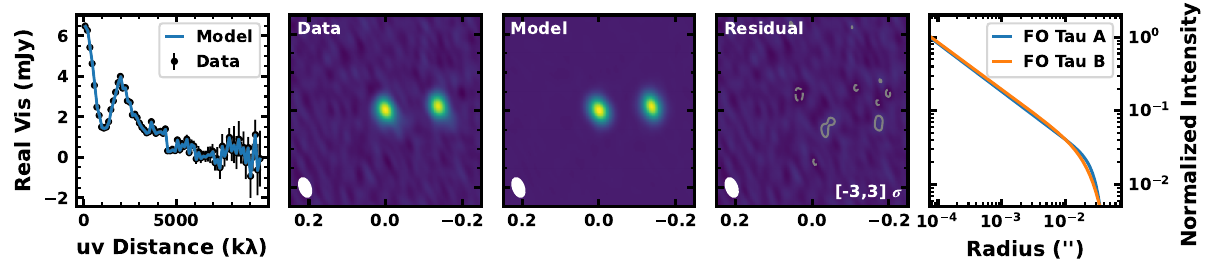}
\caption{Continuum visibility fitting. Results are shown for SPW 3. {\bf Left:} Data and model visibilities as a function of uv-distance. {\bf Middle Panels:} Cleaned images of the data, model, and residuals. Contours in the residual image are set at $-3\sigma$ and $3\sigma$ in dashed and solid lines, respectively. {\bf Right:} Best fit radial intensity profile of the primary (blue) and secondary (orange).}
\label{fig:cont_uv}
\end{center}
\end{figure*}

The results of our continuum visibility fitting are presented here. As described in Section \ref{continuum}, the four spectral windows are fit separately. The adopted value is the average of the fit values, weighted by their respective band widths. Table \ref{tabap:cont} presents the best fit model parameters for each spectral window as well as the adopted values. Figure \ref{fig:cont_uv} presents diagnostic plots for our fitting procedure, specifically for continuum SPW 3. It includes a panel or the observed and model visibilities, maps of the data, model and residuals, and the intensity profiles of the best fit models. 

\begin{deluxetable*}{l c c c c c c c c c c c}
\tablecaption{FO Tau Continuum Disk Parameters
\label{tabap:cont}}
\tablewidth{0pt}
\tabletypesize{\footnotesize}
\tablecolumns{12}
\phd
\tablehead{
  \colhead{Data Set} &
  \colhead{$F_\nu$} &
  \multicolumn{2}{c}{$R_{\rm eff,\ 68}$} &
  \multicolumn{2}{c}{$R_{\rm eff,\ 95}$} &
  \multicolumn{2}{c}{$R_{c}$} &
  \colhead{$\gamma_1$} &
  \colhead{$\gamma_2$} &
  \colhead{$i$} &
  \colhead{PA} \\
  \colhead{} &
  \colhead{(mJy)} &
  \colhead{(\arcsec)} &
  \colhead{(AU)} &
  \colhead{(\arcsec)} &
  \colhead{(AU)} &
  \colhead{(\arcsec)} &
  \colhead{(AU)} &
  \colhead{} &
  \colhead{} &
  \colhead{($^\circ$)} &
  \colhead{($^\circ$)} 
}
\startdata
\multicolumn{12}{c}{\bf FO Tau A} \\
SPW 1 & $2.9$ & $0.013$ & $1.7$ & $0.029$ & $3.9$ & $0.029$ & $4.0$ & $0.5$ & $3.2$ & $27.7$ & $121$ \\
SPW 2 & $3.2$ & $0.008$ & $1.1$ & $0.027$ & $3.7$ & $0.034$ & $4.6$ & $0.7$ & $4.6$ & $27.7$ & $120$ \\
SPW 3 & $3.2$ & $0.010$ & $1.3$ & $0.027$ & $3.7$ & $0.032$ & $4.3$ & $0.7$ & $4.2$ & $27.0$ & $122$ \\
CO SPW & $2.9$ & $0.009$ & $1.3$ & $0.027$ & $3.7$ & $0.031$ & $4.2$ & $0.6$ & $3.2$ & $26.4$ & $121$ \\
\hline
Adopted & $3.1 \pm 0.2$ & $0.010 \pm 0.002$ & $1.4 \pm 0.3$ & $0.028 \pm 0.001$ & $3.7 \pm 0.1$ & $0.032 \pm 0.002$ & $4.3 \pm 0.3$ & $0.6 \pm 0.1$ & $3.9 \pm 0.7$ & $27.3 \pm 0.5$ & $121 \pm 1$ \\
\hline
\\
\multicolumn{12}{c}{\bf FO Tau B} \\
SPW 1 & $2.8$ & $0.003$ & $0.4$ & $0.022$ & $3.0$ & $0.033$ & $4.4$ & $0.9$ & $2.2$ & $25$ & $120$ \\
SPW 2 & $3.1$ & $0.011$ & $1.4$ & $0.030$ & $4.1$ & $0.021$ & $2.8$ & $0.4$ & $1.4$ & $25$ & $121$ \\
SPW 3 & $3.1$ & $0.009$ & $1.2$ & $0.028$ & $3.8$ & $0.029$ & $3.9$ & $0.6$ & $2.1$ & $28$ & $124$ \\
CO SPW & $2.8$ & $0.007$ & $1.0$ & $0.028$ & $3.7$ & $0.016$ & $2.2$ & $0.5$ & $1.1$ & $27$ & $120$ \\
\hline
Adopted Value & $3.0 \pm 0.2$ & $0.007 \pm 0.003$ & $1.0 \pm 0.5$ & $0.027 \pm 0.004$ & $3.6 \pm 0.5$ & $0.026 \pm 0.007$ & $3.5 \pm 0.9$ & $0.6 \pm 0.2$ & $1.8 \pm 0.5$ & $26 \pm 1$ & $121 \pm 2$ \\
\enddata
\end{deluxetable*}

\section{CO Disk Modeling and Parameter Posteriors}
\label{ap:disk}

\begin{figure*}[h!]
\begin{center}
\includegraphics[width=0.98\textwidth]{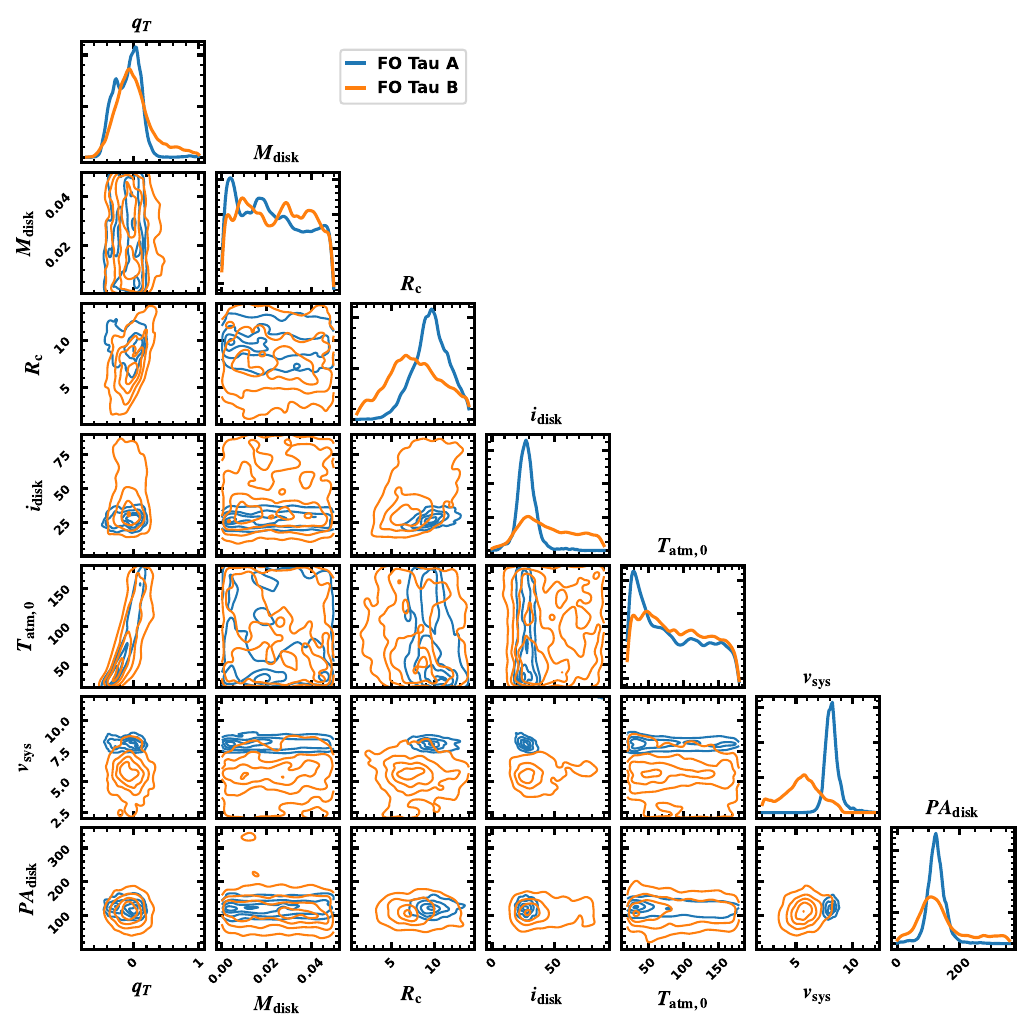}
\caption{Corner plot for parameters in the FO Tau disk fit. Posteriors and two-parameter correlation contours are represented with a Gaussian kernel-density estimate. The results for FO Tau A and B (fit simultaneously) are displayed in blue and orange, respectively.}
\label{figap:disk_corner}
\end{center}
\end{figure*}

In this appendix section we describe the structure of the disk model used in Section \ref{disk_fit}, and present the complete results of our fit. A summary of the model parameters is provided in Table \ref{tabap:disk} and a corner plot is presented in Figure \ref{figap:disk_corner}.

The disk model has a temperature structure following:
\begin{equation}
    T_{\rm gas}(r,z) = \begin{cases} T_{\rm atm}+(T_{\rm mid}-T_{\rm atm})\left(cos\frac{\pi z}{2 Z_q}\right)^{2}  & z<Z_q \\
                                T_{\rm atm}  & z\geq Z_q, 
              \end{cases}
\end{equation}
where the atmospheric temperature ($T_{\rm atm}$), mid-plane temperature ($T_{\rm mid}$), and local scale height ($Z_q$) are a function of the disk radius with the following form
\begin{equation}
    T_{\rm atm}(r) = T_{\rm atm, 0} \left(\frac{r}{150 \ AU}\right)^{q_T},
\end{equation}
\begin{equation}
    T_{\rm mid}(r) = T_{\rm mid,0} \left(\frac{r}{150 \ AU}\right)^{q_T},
\end{equation}
\begin{equation}
    Z_q(r) = Z_{q,0} \left(\frac{r}{150 \ AU}\right)^{1.3}.
\end{equation}
We adopt a simplified power-law gas surface density model of the form,
\begin{equation}
    \Sigma_{\rm gas}(r < R_c) = \frac{M_{\rm gas}(2-\gamma)}{2\pi (R_{c}^{2-\gamma}-R_{\rm in}^{2-\gamma})}r^{-\gamma},
\end{equation}
where $M_{\rm gas}$, $R_{c}$, and $R_{\rm in}$ are the disk gas mass, its outer radius, and inner calculation boundary, respectively. The disk volume density is computed assuming hydrostatic equilibrium, which in turn sets the deviation from a pure Keplerian rotation profile:
\begin{equation}
    \frac{v_{\phi}^2}{r} = \frac{GM_\star r}{(r^2+z^2)^{3/2}}+\frac{1}{\rho_{gas}}\frac{\partial P_{gas}}{\partial r}.
\end{equation}

Radiative transfer assumes local thermodynamic equilibrium with CO level populations set by the Boltzmann equation and a source function approximated by the Planck function. The emergent line shape is a Gaussian with a width of $\sqrt{2k_{\rm B}T(r,z)/m_{\rm CO}}$. The modeling package includes optional parameters for disk turbulence, a parameterized disk wind, and dust radiation, which are not utilized here.

\begin{deluxetable*}{l c c c c c}
\tablecaption{FO Tau CO Disk Parameters
\label{tabap:disk}}
\tablewidth{0pt}
\tabletypesize{\footnotesize}
\tablecolumns{6}
\phd
\tablehead{
  \colhead{Parameter} &
  \colhead{Prior} &
  \colhead{FO Tau A} &
  \colhead{FO Tau B} &
  \colhead{Description} & 
  \colhead{Note}
}
\startdata
\multicolumn{6}{l}{\textbf{Disk Structure Parameters}} \\
$q_T$                       & $\mathcal{U}(-1,1)$         & $-0.1\pm0.2$     & $-0.1\pm0.2$     & Temperature radial power law index                & Marginalized over \\
$M_{\rm disk} (M_\odot)$    & $\mathcal{U}(10^{-4},0.05$) & $0.02\pm0.02$    & $0.02\pm0.02$    & Disk Mass                                         & Marginalized over \\
$R_c$ (AU)                  & $\mathcal{U}(1,14)$         & $10\pm2$         & $8^{+4}_{-3}$    & Critical Radius                                   & Marginalized over \\
$T_{\rm atm,0}$ (K)         & $\mathcal{U}(20,180)$       & $80^{+60}_{-50}$ & $90^{+60}_{-50}$ & Atmosphere temperature normalization ($r=150$ AU) & Marginalized over \\
$\gamma$                    &                             & 1         & 1         & Surface density radial power law index                       & Fixed  \\
$R_{\rm in}$ (AU)           &                             & 1         & 1         & Inner boundary of disk density/temperature calculation       & Fixed  \\
$R_{\rm out}$ (AU)          &                             & 100       & 100       & Outer boundary of disk density/temperature calculation       & Fixed  \\
$X_{\rm CO}$                &                             & $10^{-4}$ & $10^{-4}$ & CO gas fraction relative to H$_2$                            & Fixed  \\
$Z_{q,0}$                   &                             & 33.9      & 33.9      & Vertical temperature normalization ($r=150$ AU) & Fixed  \\
$T_{\rm mid,0}$ (K)         &                             & 17.5      & 17.5      & Midplane temperature normalization ($r=150$ AU)              & Fixed  \\
$M_\star (M_\odot)$         &                             & 0.36      & 0.36      & Stellar mass                                                 & Fixed  \\
$v_{\rm turb}$ (\kms)       &                             & 0         & 0         & Turbulent velocity                                           & Fixed  \\
\hline
\multicolumn{6}{l}{\textbf{Disk Observation Parameters}} \\
$i_{\rm disk} (^\circ)$     & $\mathcal{U}(0,90)$         & $27^{+7}_{-6}$      & $40^{+30}_{-20}$   & Disk inclination     & \\
PA$_{\rm disk} (^\circ)$    & $\mathcal{U}(0,360)$        & $120^{+30}_{-20}$   & $120^{+100}_{-50}$ & Disk position angle  & \\
$v_{\rm sys}$ (\kms)        & $\mathcal{U}(2,12)$         & $8.2^{+0.6}_{-0.5}$ & $6\pm2$            & Systemic velocity    & \\
$d$ (pc)                        &                             & $135$   & $135$    & Distance                                                 & Fixed  \\
$\alpha_{\rm offset}$ (\arcsec) &                             & $0$     & $-0.137$ & Spatial offset in RA                                     & Fixed  \\
$\delta_{\rm offset}$ (\arcsec) &                             & $0$     & $0.011$  & Spatial offset in Dec                                    & Fixed  \\
$v_{\rm min}$ (\kms)            &                             & $-2.82$ & $-2.82$  & Velocity of first channel                                & Fixed  \\
$v_{\rm step}$ (\kms)           &                             & $0.635$ & $0.635$  & Channel width                                            & Fixed  \\
$n_{\rm chan}$                  &                             & $32$    & $32$     & Number of velocity channels                              & Fixed  \\
\enddata
\end{deluxetable*}

\end{document}